\documentclass[accepted]{uai2025-template}
                        
\usepackage[american]{babel}

\usepackage{natbib} 
\usepackage{mathtools} 
\usepackage{booktabs} 
\usepackage{tikz} 
\usepackage[utf8]{inputenc} 
\usepackage[T1]{fontenc}    
\usepackage{hyperref}       
\usepackage{url}            
\usepackage{booktabs}       
\usepackage{amsfonts}       
\usepackage{nicefrac}       
\usepackage{microtype}      
\usepackage{caption}
\usepackage{subfigure} 
\usepackage{multirow}
\usepackage{multicol}
\usepackage{amsmath}
\usepackage{lettrine}
\usepackage{bm}
\usepackage{algorithm}
\usepackage{algorithmicx}
\usepackage{algpseudocode}
\usepackage{graphicx}
\usepackage{textcomp}
\usepackage{multirow}
\usepackage{caption}
\usepackage{booktabs}
\usepackage{caption}
\usepackage{subfigure} 
\usepackage{threeparttablex}
\usepackage{amsthm}

\usepackage{wrapfig}
\usepackage{cancel}
\usepackage{amsmath}
\usepackage{hyperref}
\usepackage{amsfonts}
\usepackage{xcolor}
\usepackage{amsthm}
\usepackage{authblk}
\usepackage{pifont}
\definecolor{first}{RGB}{178,24,43}
\definecolor{second}{RGB}{125,115,115}
\definecolor{third}{RGB}{0,0,0}
\newtheorem{proposition}{Proposition}

\title{i$^2$VAE: Interest Information Augmentation with Variational Regularizers for Cross-Domain Sequential Recommendation}

\author[1]{Xuying Ning\thanks{Equal contribution.}}
\author[2]{Wujiang Xu$^\ast$\thanks{Correspondence to: Wujiang Xu <wujiang.xu@rutgers.edu>.}}
\author[1]{Tianxin Wei}
\author[3]{Xiaolei Liu}

\affil[1]{%
    University of Illinois Urbana-Champaign\\
    Urbana, IL, USA
}
\affil[2]{%
    Rutgers University\\
    New Brunswick, NJ, USA
}
\affil[3]{%
    Independent Researcher
}  
  
\begin{document}
\maketitle

\begin{abstract}
Cross-Domain Sequential Recommendation (CDSR) leverages user behaviors across multiple domains to mitigate data sparsity and cold-start challenges in Single-Domain Sequential Recommendation. Existing methods primarily rely on shared users (overlapping users) to learn transferable interest representations. However, these approaches have limited information propagation, benefiting mainly overlapping users and those with rich interaction histories while neglecting non-overlapping (cold-start) and long-tailed users, who constitute the majority in real-world scenarios.
To address this issue, we propose i$^2$VAE, a novel variational autoencoder (VAE)-based framework that enhances user interest learning with mutual information-based regularizers. i$^2$VAE improves recommendations for cold-start and long-tailed users while maintaining strong performance across all user groups. Specifically, cross-domain and disentangling regularizers extract transferable features for cold-start users, while a pseudo-sequence generator synthesizes interactions for long-tailed users, refined by a denoising regularizer to filter noise and preserve meaningful interest signals.
Extensive experiments demonstrate that i$^2$VAE outperforms state-of-the-art methods, underscoring its effectiveness in real-world CDSR applications.
Code and datasets are available at \url{https://github.com/WujiangXu/IM-VAE}.


\end{abstract}

\section{Introduction}
With the rise of various sequential models, single-domain sequential recommendation (SDSR) \citep{hidasi2015session, kang2018self, sun2019bert4rec, tang2018personalized, wang2020next, xu2024towards, xu2024slmrec, wei2024towards} has gained increased attention due to its ability to model users' dynamic interests in recommendation systems. However, these SDSR models often suffer from the long-standing data sparsity problem \citep{lin2021task,xu2023neural,wei2022comprehensive}, where users have few interactions in a domain to learn their preferences effectively. To address this issue, cross-domain sequential recommendation (CDSR) methods \citep{ma2019pi,ma2022mixed,cao2022contrastive,li2021dual} have been proposed to leverage abundant data from other relevant domains to improve recommendation performance in a data-scarce domain.

\begin{figure}[tb!]
\centering{
\includegraphics[width=0.98\linewidth]{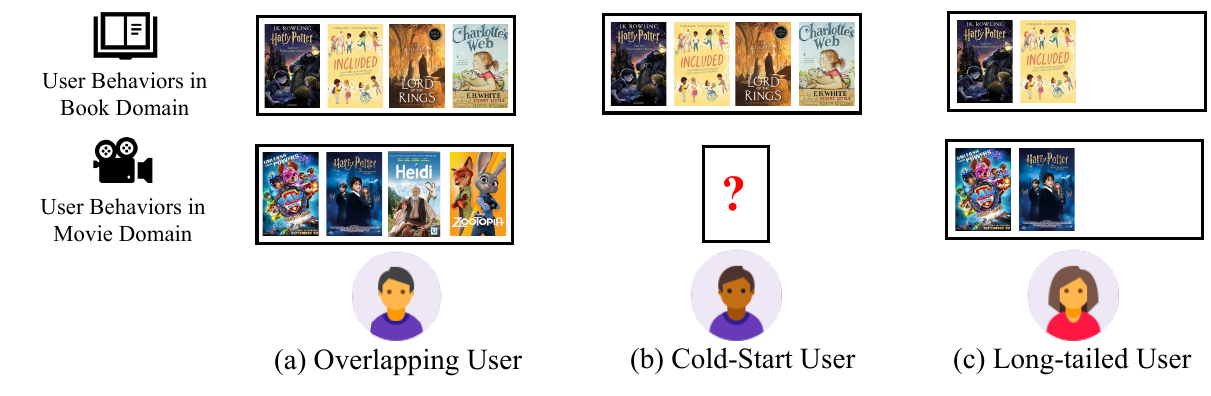}}
\caption{
Previous CDSR methods~\citep{li2021dual,ma2019pi,cao2022contrastive} rely on domain-shared information from overlapping users~(Fig.~(a)) but fail to effectively model cold-start users' interests~(Fig.~(b)) and capture long-tail users' preferences with sparse interactions~(Fig.~(c)). In contrast, i$^2$VAE disentangles cross-domain interests and refines intra-domain interest representations, enhancing recommendations for all user groups.}  
\label{Fig_infra}
\end{figure}


The core idea of CDSR is to extend SDSR methods by designing cross-domain information transfer modules to capture domain-shared interests that can be transferred across domains. Researchers have employed various techniques, including gating mechanisms~\citep{ma2019pi, sun2021parallel}, attention mechanisms~\citep{li2021dual, xu2023towards}, graph neural networks~\citep{guo2021gcn,ma2022mixed}, and contrastive learning~\citep{cao2022contrastive,xu2023towards}, to learn user interests that can be transferred across domains. 
Although it appears promising, we find that previous CDSR methods \citep{ma2019pi,li2021dual, man2017cross,cao2022contrastive} heavily rely on overlapping users (Figure 1(a)) with rich historical behaviors. They can only effectively learn users' cross-domain interests when a majority (over 70\%, as noted by \citep{xu2023rethinking}) of users are overlapping across domains, and only adequately capture intra-domain interests when users have extensive historical interactions. However, these conditions are rarely met in real-world recommendation tasks~\citep{xu2023rethinking, xu2023towards}, where the majority of users are cold-start\footnote{Cold-start users have interactions in only one domain.} (Figure 1(b)) or long-tailed\footnote{Long-tailed users are those with a few interaction records.} (Figure 1(c)). For instance, on platforms like Taobao or Amazon, there are very few overlapping users compared to the entire user base (including overlapping and non-overlapping users), and most users have very few interaction records (a.k.a. long-tailed users). In such cases, these CDSR methods, trained on the sparse historical behaviors of a few overlapping users, often show poor generalization, especially when inferring the intra-domain and cross-domain interests of cold-start and long-tailed users. This presents the \textbf{primary challenge}: how to enhance the CDSR model's performance in real-world recommendation scenarios, where most users are either cold-start or long-tailed?

Suffering from cold-start users, previous CDSR and CDR works have often utilized cross-domain modules that learn a mapping function from one domain to another based on the historical behaviors of overlapping users \citep{kang2019semi,zhu2021transfer,salah2021towards,shi2019cross}. 
Most recently, some researchers~\citep{xu2023neural,xu2023rethinking} construct user-user graph to propagate the cross-domain information for the cold-start users. However, cross-domain information is only learned in domains where users exhibit behaviors, and the density of the constructed graph heavily depends on the density of these user interactions.
Therefore, the \textbf{$\bm{1^{\text{st}}}$ sub-challenge} is: \textit{how to transfer relevant cross-domain information for cold-start users who lack historical interest data in a domain?} In addition, to address insufficient interest learning of the predominant presence of long-tailed users in real-world scenarios, MACD~\citep{xu2023towards} employs attention mechanisms to discern latent interests from users' auxiliary sequential behavior data. However, such auxiliary behaviors may not always be available in real-world scenarios. Therefore, the \textbf{$\bm{2^{\text{nd}}}$ sub-challenge} is: \textit{how to unearth and leverage the latent interest information of long-tailed users, thereby improving model performance in practical CDSR scenarios?}
To address these challenges, we propose Interest Information Augmentation with Variational Regularizers, named as i$^2$VAE, a novel framework that integrates a pseudo-sequence generator, variational autoencoders, and interest-enhancing regularizers (cross-domain, disentangling, and denoising). Our framework effectively explores latent interest information for long-tailed users while transferring disentangled cross-domain interest information for both overlapping and cold-start users, significantly improving performance in real-world scenarios.
Recent works~\citep{cao2022disencdr,cao2022cross} have explored cross-domain interest transfer but remain limited in scope. DisenCDR~\citep{cao2022disencdr} assumes full user overlap and applies mutual information for disentanglement, making it ineffective for cold-start users. CDRIB~\citep{cao2022cross} leverages information bottleneck and contrastive learning but struggles to capture long-tailed users' preferences. In contrast, our framework integrates information augmentation and denoising via mutual information, enabling robust learning for both cold-start and long-tailed users in practical scenarios.
The main contributions of our work can be summarized as follows:

\noindent$\bullet$ \quad \textbf{Novel Framework.} We propose i$^2$VAE, a framework based on variational autoencoders that enhances Interest Information through variational regularizers. It uncovers latent interests in long-tailed users and facilitates cross-domain interest transfer for overlapping and cold-start users, significantly improving real-world recommendation performance.

\noindent$\bullet$  \quad \textbf{Cold-Start Adaptation.} We design new data pathways and introduce cross-domain and disentangling regularizers to jointly model and separate users' intra- and cross-domain interests in partially overlapped CDSR scenarios.

\noindent$\bullet$  \quad \textbf{Long-Tail Interest Enhancement.} We introduce a pseudo-sequence generator combined with a denoising regularizer to enrich sparse interaction histories while eliminating noises in pseudo-sequences.
    
\noindent$\bullet$  \quad \textbf{Empirical and theoretical validation.} We show that i$^2$VAE achieves state-of-the-art performance across all user types, including long-tailed and cold-start users, in real-world cross-domain scenarios. Additionally, we provide rigorous theoretical derivations to support its effectiveness.

\vspace{-2mm}
\section{Preliminary}
This section introduces the CDSR problem, where the model uses user behavior from two domains to predict true interests. We also present the concept of mutual information.
\vspace{-3mm}
\subsection{Problem Formulation}
In this work, we consider a real-world CDSR scenario that includes a fraction of overlapping users, and a majority of long-tailed and cold-start users across two domains, namely domain $X$ and domain $Y$. The recommendation data is represented by {\small $D^X=(\mathcal{U}^X,\mathcal{V}^X,\mathcal{E}^X)$} and {\small $D^Y=(\mathcal{U}^Y,\mathcal{V}^Y,\mathcal{E}^Y)$}, where {\small $\mathcal{U}^{\cdot}$}, {\small $\mathcal{V}^{\cdot}$}, and {\small $\mathcal{E}^{\cdot}$} are the sets of users, items, and interaction edges, respectively. For a given user, we denote the sequeneces of user-item interaction in chronological order, as {\small $S^X=[v^X_1,v^X_2,\cdots,v^X_{|S^X|}]$} and {\small $S^Y=[v^Y_1,v^Y_2,\cdots,v^Y_{|S^Y|}]$}, where {\small $|\cdot|$} is the length of user behavior. The objective of CDSR is to predict the next item that each user will purchase based on the user's previous behavior in two domains~\citep{cao2022cross,xu2023towards}.

\subsection{Preliminary of Mutual Information}
Mutual Information Maximization~\citep{mcgill1954multivariate,hjelm2018learning,belghazi2018mutual} is a key mathematical tool for ensuring the robustness of interest augmentation. For random variables \(X\) and \(Y\), the \textit{mutual information}~\citep{jarvelin2002cumulated,paninski2003estimation} \(I(X; Y)\), measuring how much \(X\) reduces uncertainty in \(Y\), is defined as: \(I(X; Y) = H(X) - H(X\vert Y)\),
where \(H(X)\) and \(H(X\vert Y)\) denote the entropy~\citep{renyi1961measures,cover1991entropy} of \(X\) and the conditional entropy of \(X\) given \(Y\). For three random variables \(X\), \(Y\), and \(Z\), the \textit{interaction information}~\citep{mcgill1954multivariate,bell2003co} \(I(X; Y; Z)\), is defined as:
\begin{equation}
    I(X; Y; Z) = I(X; Y) - I(X; Y \vert Z), \label{interaction information}
    \vspace{-1mm}
\end{equation}
which captures shared information beyond pairwise mutual information. Notably, all variables are symmetric in these definitions.

\vspace{-3mm}
\section{Methodolodgy}
\begin{figure*}[tb!]
    \centering
    \includegraphics[width=0.9\textwidth]{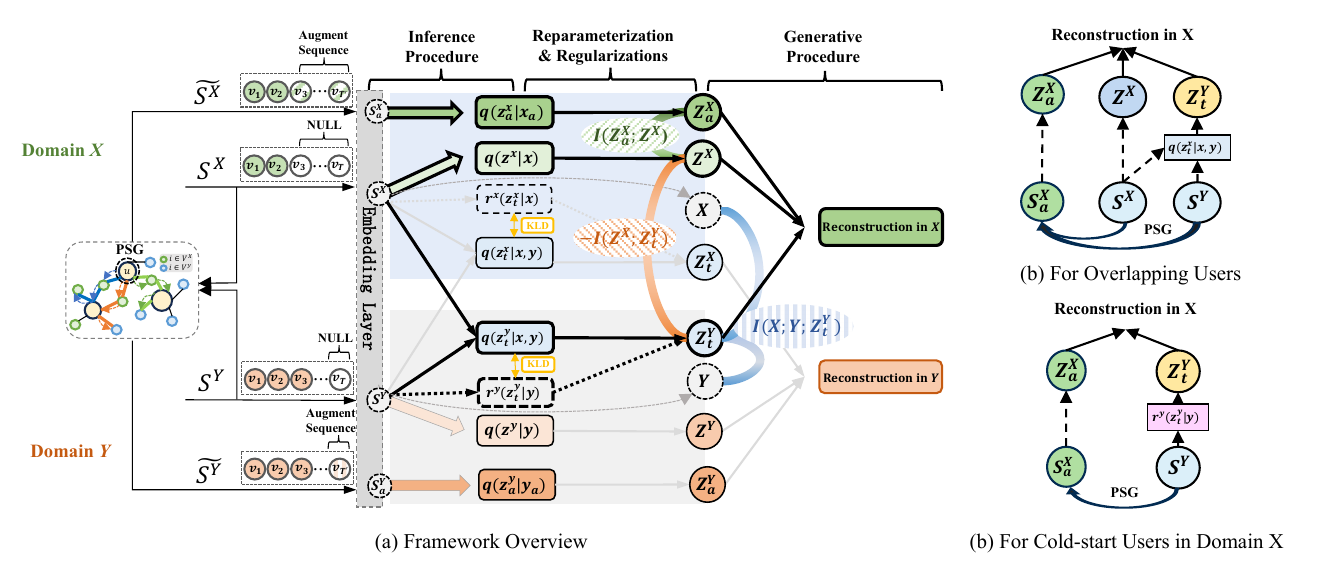}
    \vspace{-8pt}
    \caption{Figure (a) presents an overview of i$^2$VAE, highlighting computational pathways (thick black lines) and key interest-enhancing regularizers {\small ${I(\bm{Z^X_a}; \bm{Z^X})}$}, {\small $-I(\bm{Z^X};\bm{Z^Y_t})$}, and {\small $I(\bm{X};\bm{Y};\bm{Z^Y_t})$} for predictions in domain $X$. The dashed component $r^y(z^y_t|y)$ replaces the cross-domain module $q(z^y_t|x,y)$ for cold-start users in $X$. The model is symmetric for domain $Y$ (grey lines). Figures (b) and (c) depict simplified data pathways for overlapping and cold-start users in domain $X$.}
    \vspace{-12pt}
    \label{framework}
\end{figure*}

This section introduces the pseudo-sequence generator, inference and generation procedures of i$^2$VAE, and its interest-enhancing regularizers. Figure \ref{framework} provides an overview.

\vspace{-3mm}
\subsection{Pseudo-Sequence Generator}

Behavior sparsity poses a significant challenge in real-world CDSR scenarios, making it difficult for models to capture users' within-domain interests, let alone cross-domain interests. Thus, the pseudo-sequence generator (PSG) \textit{augments user behaviors} by serving as a fast retrieval model that efficiently generates candidate items aligned with user interests but not yet interacted with. To balance efficiency and effectiveness, we use LightGCN~\citep{he2020lightgcn} with iterative recall to generate pseudo-sequences while avoiding the high computational cost of sequential recommendation models.

The pseudo-sequence generation process consists of three steps: (1) Unified Item Set Construction: Items from both domains are remapped into a unified item set, {\small \( \mathcal{V} = \mathcal{V}^X \cup \mathcal{V}^Y \)}, ensuring no duplicates. A user-item bipartite graph {\small \( \mathcal{G} \)} is built based on the interaction data {\small \( \mathcal{E} = \mathcal{E}^X \cup \mathcal{E}^Y \)}. (2) Embedding Learning \& Recall: Using LightGCN, user and item embeddings are learned as {\small \( \bm{E}_\mathcal{U} \in \mathbb{R}^{|\mathcal{U}| \times d} \)} and {\small \( \bm{E}_\mathcal{V} \in \mathbb{R}^{|\mathcal{V}| \times d} \)}, where {\small \( d \)} is the embedding dimension. Predicted user preferences are represented by the rating matrix {\small \( \bm{R} = \bm{E}_\mathcal{U}\bm{E}_\mathcal{V}^T \)}. The recall process iteratively selects candidate items based on {\small \( \bm{R} \)},  updates the interaction graph with pseudo-interactions, and refines embeddings through message propagation. (3) Pseudo-Sequence Generation: Users' original interaction sequences {\small \( S^{X} \)} and {\small \( S^{Y} \)} are expanded with high-scoring items from the recall process that have not been interacted with. For domain \( X \), the pseudo-sequence is defined as  \( \widetilde{S}^{X} = S^{X} \cup \{ \widetilde{v}_1, \ldots, \widetilde{v}_{T'} \}, \)   
where the top {\small \( T' \)} items {\small \( \widetilde{v}_i \)} are selected by the iterative recall process, satisfying  \( \widetilde{v}_i \in \mathcal{V}^X \setminus S^{X}\) . A similar process is applied to domain \( Y \).

\subsection{Embedding Layer}
We employ embedding layers {\small$\bm{E}^X \in \mathbb{R}^{|\mathcal{V}^X| \times d}$} and {\small$\bm{E}^Y \in \mathbb{R}^{|\mathcal{V}^Y| \times d}$}, along with the self-attention layer from SASRec~\citep{kang2018self}, to derive interest representations for both \textit{real} sequences ({\small \( S^{X} \)} and {\small \( S^{Y} \)}) and \textit{pseudo} sequences ({\small{\( \widetilde{S}^{X} \)}} and {\small{\( \widetilde{S}^{Y} \)}}). 
Specifically, the representations are computed as  
{\small  
\(
\bm{S^X}= \odot \big( \{\mathbf{h}_{S_{1}^{X}},\cdots ,\mathbf{h}_{S_{T}^{X}}\}\big),\) 
\(\bm{S^Y}= \odot\big( \{\mathbf{h}_{S_{1}^{Y}},\cdots,\mathbf{h}_{S_{T}^{Y}}\} \big),
\)
\(
\bm{S^X_a}=\odot \big( \{\mathbf{h}_{\widetilde{S}_1^X},\cdots,\mathbf{h}_{\widetilde{S}_{T'}^X}\} \big),  \)
\( \bm{S^Y_a}=\odot \big( \{\mathbf{h}_{\widetilde{S}_1^Y},\cdots,\mathbf{h}_{\widetilde{S}_{T'}^Y}\} \big),
\)
}
where \( \mathbf{h}_{S_t^X} \) and \( \mathbf{h}_{S_t^Y} \) denote the hidden representations of items at position \( t \) in sequences \( S^X \) and \( S^Y \), and \( \odot \) represents mean pooling over the time dimension, producing embeddings in {\small $\mathbb{R}^d$}.
Here, \( T \) and \( T' \) are the maximum lengths of the real and pseudo sequences, respectively. For sequences longer than \( T \) (\( T' \)), only the most recent \( T \) (\( T' \)) actions are retained; for shorter sequences, 'padding' items are added to the left to meet the required length. {\small $\bm{S^X}$} and {\small $\bm{S^Y}$} represent the user interests derived from real sequences, while {\small $\bm{S^X_a}$} and {\small $\bm{S^Y_a}$} capture the \textit{augmented interests }obtained from pseudo sequences.

\vspace{-2mm}
\subsection{Generation and Inference of i$^2$VAE}
In line with previous CDR studies~\citep{cao2022disencdr, cao2022contrastive}, we adopt variational autoencoder as the foundamental architecture due to its ability to model interest decomposition and transfer. We assume that a user's interests in domains \( X \) and \( Y \) are represented by {\small \( \bm{X} \)} and {\small \( \bm{Y} \)}, which capture the user's true preferences in each domain. These interests follow a bivariate distribution {\small \( P_D(\bm{X}, \bm{Y}) \)}, and our goal is to reconstruct them in {\small \( P_D \)} to predict the next items the user will interact with. In our i$^2$VAE, to map the initial interest representations (\( \bm{S^X}, \bm{S^Y}, \bm{S^X_a}, \bm{S^Y_a} \)) to their true interests, we design six \( d \)-dimensional latent variables in VAE, three for each domain, representing different aspects of user interests to facilitate interest reconstruction:

$\bullet$ \( \bm{Z^X} \), \( \bm{Z^Y} \): \textit{Domain-specific interests} in domain \( X \) and \( Y \). 

$\bullet$ \( \bm{Z^Y_t} \), \( \bm{Z^X_t} \): \textit{Transferable interests} that capture cross-domain user preferences; \( \bm{Z^Y_t} \) represents the interests in domain \(Y\) that are transferable to \(X\), and vise versa for \( \bm{Z^X_t} \).

$\bullet$ \( \bm{Z^X_a} \), \( \bm{Z^Y_a} \): \textit{Augmented domain-specific interests} derived from pseudo sequnces in domain \( X \) and \( Y \).

\vspace{-3mm}
\subsubsection{\textbf{Inference Procedure}}
\label{subsec:inference-procedure}
To model the six representations described above, we use the VAE~\citep{kingma2013auto}, which encodes different aspects of a user's interests into latent variables. We assume that the six latent variables are conditionally independent given {\small \( \bm{X} \)} and {\small \( \bm{Y} \)}. Moreover, {\small $\bm{Z_t^Y}$} and $\bm{Z_t^X}$ represent cross-domain information, while {\small $\bm{Z^X}$}, {\small $\bm{Z_a^X}$}, {\small $\bm{Z^Y}$}, and {\small $\bm{Z_a^Y}$} represent domain-specific information that only correlates with one domain. 
With the previously mentioned assumptions, we can obtain {\small$q(z^x \vert x, y) = q(z^x \vert x)$}\footnote{To keep equations compact, we use lowercase, non-bold symbols in prior and posterior expansions to represent their bold uppercase counterparts (e.g., \( z_t^y \) for \( \bm{Z_t^Y} \), and \( x \) for \( \bm{X} \)).}, which is applicable to other domain-specific latent variables. Based on this assumption, we factorize $q_\phi$ as follows.
\begin{small}
\begin{align}
    &\quad q(z^x, z^y, z_t^y, z_t^x, z_a^x, z_a^y \vert x, y) \label{inference q}  \\
    &= \underbrace{q(z^x \vert x) q(z_t^y \vert x, y) q(z_a^x \vert x)}_{\textit{Domain } X} \underbrace{q(z^y \vert y) q(z_t^x \vert x, y) q(z_a^y \vert y)}_{\textit{Domain } Y} \nonumber
\end{align}
\end{small}
Resembling standard VAE's inference process, each \( q(\cdot) \) in the factorization above corresponds to a VAE's encoder, and it is assumed to follow a Gaussian distribution. The mean (\( \mu^{\cdot} \)) and standard deviation vectors (\( \sigma^{\cdot} \)) of the distribution are generated by their respective encoders. In the equation above, {\small \( \phi \)} represents the learnable parameters of all encoders. Details regarding the implementation of domain-specific and cross-domain encoders are provided in Appendix~\ref{sec:implement_encoder_HP}. Then, using the reparameterization trick, latent variables are generated as follows. For \( z^x \), we have:
\begin{small}
\begin{equation}
     z^x = \mu^x + \sigma^x \odot \epsilon, \ \ \epsilon \sim \mathcal{N}\left(\bm{0}, \bm{I}\right),
\end{equation}
\end{small}
with other latent variables generated similarly.
\vspace{-3mm}
\subsubsection{\textbf{Generative Procedure}}
We assume that {$\bm{X}$} and { $\small\bm{Y}$} are conditionally independent given the six latent variables. Additionally, each domain is associated with only three latent variables. For instance, reconstructing {\small $\bm{X}$} only involves the domain-specific interest representation {\small $\bm{Z^X}$}, the augmented interest representation {\small $\bm{Z^X_a}$}, and the cross-domain representation {\small $\bm{Z^Y_t}$} for information transferred from domain $Y$ to $X$. Therefore, we have {\small$p_{\theta}(x|z^x, z^y_t, z^x_a, z^y, z^x_t, z^y_a) = p_{\theta_{x}}(x|z^x, z^y_t, z^x_a)$}, which similarly applies to domain $Y$.
Based on this assumption, our generative distribution can be structured as:
\vspace{-2mm}
\begin{equation}
\resizebox{0.9\columnwidth}{!}{$
\begin{aligned}
    p_\theta(x,y) = &  \int p_{\theta_x}(x|z^x,z^y_t,z^x_a) p_{\theta_y}(y|z^y,z^y_t,z^y_a)p(z^x) p(z^y) \nonumber \\
    &\quad \cdot p(z^x_a){p}(z^y_a) p(z^x_t) p(z^y_t) dz^x dz^y dz^x_a dz^y_a dz^x_t dz^y_t.  
\end{aligned}
$}
\end{equation}
Here, \( \theta = \{\theta_x, \theta_y\} \) denotes the parameters of the VAE decoders. The prior distributions $p(\cdot)$ for cross-domain latent variables (\( z^x_t \) and \( z^y_t \)) are set as standard normal distributions, \( \mathcal{N}(\bm{0}, \bm{I}) \), while domain-specific priors vary in mean and standard deviation to reflect domain distinctions.
\vspace{-3mm}
\subsubsection{\textbf{Evidence Lower Bound of i$^2$VAE}} 
Similar to standard VAE approaches, we derive the Evidence Lower Bound (ELBO) of the optimization objective~\citep{kingma2013auto, salah2021towards} based on the assumptions in the generation and inference. It comprises two reconstruction terms ($\mathbb{E}_{q_{\phi}}[\cdot]$), aiming to reconstruct the user's true interests in domains \(X\) and \(Y\), and six Kullback-Leibler (KL) divergences ($D_{KL}[\cdot]$), which regularize the latent variable distributions to align with their priors:
\begin{equation}
\resizebox{0.9\columnwidth}{!}{$
\begin{aligned}
\label{ELBO}
    \log p(x,y)  & \ge  \text{ELBO}  \\
    & = \mathbb{E}_{q_{\phi}}[\log p(x\vert{z}^x,z_t^y,z_a^x)] +\mathbb{E}_{q_{\phi}}[\log p(y\vert z^y,z_t^x,z_a^y)] \\
    &\  -D_{KL}[q(z^{x}\vert x)\Vert p(z^{x})]-D_{KL}[q(z^{y}\vert y)\Vert p(z^{y})]  \\ 
    &\  -D_{KL} [q(z_t^y|x,y)\Vert p(z_{c}^{y})]
    -D_{KL}[q(z_t^x|x,y)\Vert p(z_{c}^{x})]  \\
    &\  -D_{KL}[q(z_a^x|x)\Vert p(z_{a}^{x})]
    -D_{KL}[q(z_a^y|y)\Vert p(z_{a}^{y})].
\end{aligned}
$}
\end{equation}
For the reconstruction term in domain \( X \), an MLP functions as the VAE decoder, reconstructing the user's true interest representation \( \bm{X} \in \mathbb{R}^d \) from three domain-specific latent variables. Each item in domain \( X \) is associated with an embedding \( \bm{E}_{v_i}^X \in \mathbb{R}^d \), which is used to estimate the user's preference for an item via their dot product:  
\(\hat{r}_{ui}^X = \bm{X}_u \cdot \bm{E}_{v_i}^X.\)
The reconstruction term is optimized by minimizing the Binary Cross-Entropy Loss between the predicted interaction probability \( \hat{r}_{ui}^X \) and the true interaction label \( r_{ui}^X \). The same process applies to domain \( Y \).


\subsection{Interest-Enhancing Regularizers}
Using an unconstrained VAE model alone cannot guarantee reliable interest augmentation. We identify two key aspects of this reliability: (a) Predicting interactions in the cold-start domain relies \textit{solely} on cross-domain transferable interests, which must be \textit{separate} from domain-specific interests in the other domain. (b)~Augmenting intra-domain interests may introduce noise due to inaccuracies in the PSG, thus a \textit{denoising mechanism} is needed to filter out these deviations and preserve the true intra-domain augmented interests. To address these reliability challenges, we propose three interest-enhancing regularizers—cross-domain, disentangling, and denoising—illustrated using domain \( X \), with symmetric designs for domain \( Y \) omitted for brevity.


\vspace{-2mm}
\subsubsection{\textbf{Cross-domain Regularizer \texorpdfstring{\( I(\bm{X};\bm{Y};\bm{Z_t^Y}) \)}{I(X;Y;Zc^Y)}}}
We aim for the the transferable interest representation \( \bm{Z_t^Y} \) to capture user interests that can transfer from domain \( Y \) to \( X \), acting as a cross-domain signal, such as \textit{shared themes or topics across domains like books and movies.} To achieve this, we propose the first regularizer—maximizing the interaction information \( I(\bm{X}; \bm{Y}; \bm{Z_t^Y}) \)—which quantifies the interdependence and shared information among these variables. Using the formal definition of interaction information (Eq.~\eqref{interaction information}), we expand \( I(\bm{X};\bm{Y};\bm{Z_t^Y}) \) as:
\begin{small}
\begin{align}
I(\bm{X};\bm{Y};\bm{Z_t^Y}) = I(\bm{X};\bm{Z_t^Y}) - I(\bm{X};\bm{Z_t^Y} \vert \bm{Y}). \label{I(X;Y;Z_t^y)}
\end{align}
\end{small}
The above equation can be intuitively explained from two perspectives: (1) it increases \( I(\bm{X};\bm{Z_t^Y}) \), ensuring \( \bm{Z_t^Y} \) captures information relevant to \( \bm{X} \); (2) it minimizes \( I(\bm{X};\bm{Z_t^Y} \vert \bm{Y}) \), which helps restrict \( \bm{Z_t^Y} \) to information inferred from domain \( Y \). By doing so, the cross-domain latent variable, \( \bm{Z_t^Y} \), effectively encodes shared and transferable user-interest signals across the two domains.

\vspace{-2mm}
\subsubsection{\textbf{Disentangling Regularizer \texorpdfstring{\(-I(\bm{Z^X}; \bm{Z_t^Y})\)}{-I(Z^X; Zc^Y)}}}

For cold-start users in domain \( X \), only \( \bm{Z_t^Y} \), the cross-domain representation, is utilized for reconstruction, while all domain-specific representations in \( X \) are masked. This design requires \( \bm{Z_t^Y} \) to focus \textit{exclusively} on transferable cross-domain interests without contamination from domain-specific information in \( Y \), as such contamination could lead to negative transfer~\citep{zhang2023collaborative,park2023cracking}.

To achieve this, we propose our second regularizer-minimizing {\small \( I(\bm{Z^X}; \bm{Z_t^Y}) \)}-ensuring that {\small \( \bm{Z_t^Y} \)} remains disentangled and dedicated to cross-domain information. We further propose \textit{a generalizable proposition} that simplifies intractable disentangling regularizers by breaking them down into manageable components. Distinct from \citet{cao2022disencdr,cao2022contrastive}, our \textit{derivation of its optimizable terms} in Section~\ref{sec:interest reg} further ensures the reliability of interest extraction for cold-start users.

\begin{proposition}
Let \( \bm{Z^X} \in \mathbb{R}^d \) represent the domain-specific interest in domain \( X \), and \( \bm{Z_t^Y} \in \mathbb{R}^d \) denote the cross-domain transferable interest from domain \( Y \) to \( X \). To effectively disentangle domain-specific and cross-domain interests, we aim to minimize their mutual information, \( I(\bm{Z^X}; \bm{Z_t^Y}) \), which is equivalent to:
\vspace{-1mm}
\begin{small}
\begin{equation}
    \max \Big \{ -I(\bm{X}; \bm{Z^X}) - I(\bm{X}; \bm{Z_t^Y}) + I(\bm{X}; \bm{Z^X}, \bm{Z_t^Y}) \Big \}. \label{theory_1}
\end{equation}
\end{small}
\end{proposition}
\vspace{-2mm}
The proof is given in the Appendix~\ref{sec:theory_r1}. Eq.~\eqref{theory_1} above can be intuitively explained as follows: $\bm{Z^X}$ and $\bm{Z^Y_t}$ are required to be jointly informative to domain $X$ (the third term), and the total amount of information in $\bm{Z^Y_t}$ and $\bm{Z^X}$ will be penalized (the first and second terms). Thus, maximizing Eq.~\eqref{theory_1} will naturally encourage $\bm{Z^Y_t}$ and $\bm{Z^X}$ to encode the distinct, non-overlapping information that can be informative to domain $X$. 
\vspace{-3mm}
\subsubsection{\textbf{Denoising Regularizer \texorpdfstring{\( I(\bm{Z^X}; \bm{Z^X_a}) \)}{I(Z^X; Z^X_a)}}}
\label{sec:Denoising_Regularizer}
The third regularizer aims to maximize the mutual information \( I(\bm{Z^X}; \bm{Z^X_a}) \) between the representations of \( S^X \) and \( \widetilde{S}^X \). This ensures that the pseudo-sequences retain relevant intra-domain interest information while filtering out noise introduced by unreliable recalled items from PSG. Specifically, maximizing \( I(\bm{Z^X}; \bm{Z^X_a}) \) reduces the uncertainty of \( \bm{Z^X} \) given \( \bm{Z^X_a} \), encouraging \( \bm{Z^X_a} \) to capture the true underlying information in \( \bm{Z^X} \) and eliminate irrelevant noise.
When user behavior is sparse, it is crucial to leverage rich information from the pseudo-sequence rather than excessively denoise it. Conversely, when user behavior is abundant, enhancing the denoising process improves the model's ability to capture useful augmentations. Therefore, we introduce a noise-adaptive weight \(\lambda_{d}^X\) for the denoising regularizer \(I(\bm{Z^X}; \bm{Z^X_a})\), which increases with the richness of user interactions: \( \lambda_{d}^X = \text{exp}(aL^X/T) - b \), where \(L^X\) is the length of the user's unpadded historical behavior in domain \(X\), and \(T\) is the maximum length of \( S^X \). Constants \(a\) and \(b\) both set to 0.8 in our study.

\vspace{-2mm}
\subsection{Optimization Details} 
\subsubsection{Derivation of Regularizers' Lower Bounds} 
\label{sec:interest reg}
Combining the previous three interest-enhancing regularizers for domain $X$, we have the following optimization objectives to enforce robustness:
\begin{equation}
\resizebox{0.85\columnwidth}{!}{
    $\max \Bigg\{\underbrace{ 
    \overbrace{I(\bm{X};\bm{Y};\bm{Z_t^Y})}^{\text{Cross-domain}} 
    +
    [\overbrace{-I(\bm{Z^X}; \bm{Z_t^Y})}^{\text{Disentangling}}] 
    + 
    \overbrace{I(\bm{Z^X};\bm{Z^X_a})}^{\text{Denoising}} 
    }_{I_{\text{interest}}^X}\Bigg\},$
} \label{maximize_obj2}
\end{equation}

with the derivations in Eq.~\eqref{I(X;Y;Z_t^y)} - \eqref{theory_1}, it is equivalent to:
\begin{equation}
\resizebox{0.9\columnwidth}{!}{
    $\max \Big\{ \underbrace{ I(\bm{X}; \bm{Z^X}, \bm{Z^Y_t}) -I(\bm{X}; \bm{Z^X}) -I(\bm{X};\bm{Z_t^Y}\vert \bm{Y}) + I(\bm{Z^X}; \bm{Z^X_a}) }_{I_{\text{interest}}^X}\Big\}. $
} \label{maximize_obj}
\end{equation}


Although mutual information possesses elegant mathematical properties, the terms above cannot be directly optimized, requiring innovative derivations to \textit{establish tractable and reliable lower bounds}. Below, we provide derivations and explanations for bounds $\mathcal{L}_{(\cdot)}$.

\noindent $\bullet$ \quad \textbf{Derivations for {\small $I(\bm{X};\bm{Z^X},\bm{Z_t^Y}) \geq\mathcal{L}_{I(\bm{X}; \bm{Z^X}, \bm{Z_t^Y})}$}}:
\begin{equation}
\resizebox{0.9\columnwidth}{!}{
    $I(\bm{X};\bm{Z^X},\bm{Z_t^Y}) \geq H(\bm{X})  
    + \mathbb{E}_{p_{D}q(z^{x}\vert x)q(z_t^y\vert x,y)}
    \left[ \log p(x\vert z^{x},z_t^y)\right]$
} \label{max_t1}
\end{equation}
The detailed derivation is provided in Section~\ref{sec:Section_max_t1}. Maximizing Eq.~\eqref{max_t1} plays a critical role in aligning {\small $p(x \vert z^{x}, z_t^y)$} with {\small $q(x \vert z^{x}, z_t^y)$}, effectively functioning as a reconstruction term. This alignment ensures that the learned latent representations {\small $\bm{Z^X}$} and {\small $\bm{Z_t^Y}$} are capable of accurately reconstructing the original user preferences {\small $\bm{X}$}. To avoid redundancy in the objectives, we approximate this term in practice with {\small $\mathbb{E}_{q_{\phi}}[\log p(x\vert z^x,z_t^y,z_a^x)]$} from Eq.~\eqref{ELBO}. This approximation not only avoids overlapping objectives but also simplifies computation. For clarity, we refer to this term as {\small $\mathcal{L}_{I(\bm{X}; \bm{Z^X}, \bm{Z_t^Y})}$}.

\noindent $\bullet$ \quad  \textbf{Derivations for {\small $-I(\bm{X};\bm{Z^X})\geq\mathcal{L}_{-I(\bm{X};\bm{Z^X})}$}}:
\begin{small}
\begin{equation}
\begin{split}
-I(\bm{X};\bm{Z^X}) &= -\mathbb{E}_{p_{D}(x)}\left[\ D_\text{KL} \left[ q(z^x \vert x) \Vert q(z^x)\right] \right]  \\
& \geq -\mathbb{E}_{p_{D}(x)}\left[\ D_\text{KL} \left[ q(z^x \vert x) \Vert p(z^x)\right] \right]. \label{max_t2}
\end{split}
\end{equation}
\end{small}
This formula represents reducing the mutual information between \( \bm{X} \) and its latent representation \( \bm{Z^X} \) to ensure that \( \bm{Z^X} \) retains only the necessary information about \( \bm{X} \). By replacing \( q(z^x) \) with the prior distribution \( p(z^x) \), a more tractable KL divergence lower bound is obtained for practical model training. We denote the derived term as {\small $\mathcal{L}_{-I(\bm{X};\bm{Z^X})}$}.

\noindent $\bullet$ \quad \textbf{Derivations for {\small $-I(\bm{X};\bm{Z_t^Y}\vert \bm{Y}) \geq\mathcal{L}_{-I(\bm{X}; \bm{Z_t^Y} \mid \bm{Y})}$}}:
\begin{equation}
\resizebox{0.88\columnwidth}{!}{
    $-I(\bm{X};\bm{Z_t^Y}\vert \bm{Y}) \geq 
    -\mathbb{E}_{p_{D}(x,y)}\left[ D_{KL}\left( q(z^{y}_{t}\vert x,y) 
    \Vert  r^y(z^{y}_{t}\vert y) \right) \right]$
} \label{max_t3}
\end{equation}
The derived bound is denoted as {\small $\mathcal{L}_{-I(\bm{X}; \bm{Z_t^Y} \mid \bm{Y})}$}, with detailed derivations provided in Section~\ref{sec:Section_max_t3}. Minimizing {\small $-I(\bm{X}; \bm{Z_t^Y} \mid \bm{Y})$} aligns the auxiliary distribution {\small $r^y(z^y_t \vert y)$} with the cross-domain representation {\small $q(z^y_t \vert x, y)$}. This is particularly useful for cold-start users in domain \(X\), where {\small $q(z^y_t \vert x, y)$} becomes unreliable. In such cases, as illustrated in Figure \ref{framework}, {\small $r^y(z^y_t \vert y)$}, relying solely on domain \(Y\), substitutes {\small $q(z^y_t \vert x, y)$} during inference to ensure the robustness for cold-start users.

\noindent $\bullet$ \quad \textbf{Derivations for {\small $I(\bm{Z^X}; \bm{Z^X_a}) \ge \mathcal{L}_{I(\bm{Z^X}; \bm{Z^X_a})}$}}:
\begin{equation}
\resizebox{0.88\columnwidth}{!}{
  $ I(\bm{Z^X}; \bm{Z^X_a}) = -\mathbb{E}_{q(z^x_a\vert z^x,x)}\left[D_{KL}\left(q(z^x\vert x)\Vert q(z^x_a\vert x)\right)\right] + \epsilon. $} \label{max_denoise}
\end{equation}
Detailed derivation can be found in Section~\ref{sec:Section_max_denoise}. Since $\epsilon$ is intractable, we optimize {\small $I(\bm{Z^X};\bm{Z^X_a})$} by solely maximizing the first term, denoted as the lower bound {\small $\mathcal{L}_{I(\bm{Z^X}; \bm{Z^X_a})}$}.

\subsubsection{Overall Optimization Objectives}
The interest-enhancing regularizers for the reconstruction of domain \( Y \) are symmetrical. Referring to Eqs.~\eqref{max_t1}-\eqref{max_denoise}, we can obtain optimization objectives for domain $Y$: {\small$\mathcal{L}_{I(\bm{Y};\bm{Z^Y},\bm{Z_t^X})}$}, {\small$\mathcal{L}_{-I(\bm{Y};\bm{Z^Y})}$}, {\small$\mathcal{L}_{-I(\bm{Y};\bm{Z_t^X} \vert \bm{X})}$}, and {\small$\mathcal{L}_{I(\bm{Z^Y}; \bm{Z^Y_a})}$}.

To jointly optimize domains \( X \) and \( Y \), we derive the overall optimization objective by combining the ELBO in Eq.~\eqref{ELBO} with the tractable lower bounds of $I_{\text{interest}}^{(\cdot)}$ for each domain $X$ and $Y$. This is achieved using balancing weights \( \lambda_{a} \) and adaptive denoising weight $\lambda_{d}^{(\cdot)}$ for the denoising regularizer and \( \lambda_{c} \) for the cross-domain and disentangling regularizers. The overall optimization objective is formulated as follows.
\resizebox{0.95\columnwidth}{!}{
\begin{minipage}{\columnwidth}
\begin{align}
    \max_{\theta, \phi} \Big\{ &\log p(x,y) + I_{\text{interest}}^X + I_{\text{interest}}^Y \Big\} \notag \\
    \geq \max_{\theta, \phi}\Big\{& \text{ELBO} 
    + \lambda_{a} \lambda_{d}^X \mathcal{L}_{I(\bm{Z^X}; \bm{Z^X_a})} 
    + \lambda_{a} \lambda_{d}^Y \mathcal{L}_{I(\bm{Z^Y}; \bm{Z^Y_a})} \notag \\
    &+ \lambda_{c} \big( 
        \mathcal{L}_{I(\bm{X}; \bm{Z^X}, \bm{Z_t^Y})} 
        + \mathcal{L}_{-I(\bm{X}; \bm{Z^X})} 
        + \mathcal{L}_{-I(\bm{X}; \bm{Z_t^Y} \vert \bm{Y})} 
      \big)  \notag \\
    &+ \lambda_{c} \big( 
        \mathcal{L}_{I(\bm{Y}; \bm{Z^Y}, \bm{Z_t^X})} 
        + \mathcal{L}_{-I(\bm{Y}; \bm{Z^Y})} 
        + \mathcal{L}_{-I(\bm{Y}; \bm{Z_t^X} \vert \bm{X})}  \notag
      \big) 
    \Big\}.
\end{align}
\end{minipage}
}


\vspace{-3mm}
\section{Experiments}
In this section, we conduct experiments to evaluate the performance of our i$^2$VAE. Experiments in this section intend to answer the following research questions (RQs):
\textbf{RQ1}: How does i$^2$VAE perform compared to other baseline methods in the CDSR task across different user types, including long-tailed and cold-start users? \textbf{RQ2}: How do the different modules of i$^2$VAE contribute to the performance improvement of our method? \textbf{RQ3}: Can i$^2$VAE consistently achieve strong performance across varying user-item interaction densities and different numbers of overlapping users, and how do hyperparameter settings impact its performance?

\renewcommand{\arraystretch}{1.2}
\begin{table*}[tb!]
\small
\setlength\tabcolsep{0.5pt}
\setlength{\abovecaptionskip}{1.5pt}
\setlength{\belowcaptionskip}{2pt}
\caption{ Experimental Results ($\%$) across different types of users, including long-tailed(tailed), cold-start, and all users, on "Cloth-Sport" and "Phone-Elec" CDSR datasets. Due to space constraints, the full experimental results, including the "Game-Video" dataset, are presented in Appendix~\ref{sec:remaining_tab}. The \textcolor{first}{\textbf{best}} and \textcolor{second}{\underline{\textbf{second-best}}} average performances are highlighted.}
\label{Main Results}
\resizebox{\textwidth}{!}{
\begin{tabular}{c|c|c|ccc|ccc|ccc|c|c}
\toprule
\multirow{2}{*}{\phantom{0}Datasets \phantom{0}} & \multirow{2}{*}{\phantom{0} User Types\phantom{0}} & \multirow{2}{*}{\phantom{0} Metric\phantom{0}} & \multicolumn{3}{c|}{SDR\phantom{0}} & \multicolumn{3}{c|}{CDR-sequential\phantom{0}} & \multicolumn{3}{c|}{CDR\phantom{0}} & \multicolumn{1}{c|}{Ours\phantom{0}} & \multirow{2}{*}{\phantom{0}\textuparrow ($\%$)}\\ \cline{4-13}
   & & & Multi-VAE  & SVAE & SASRec  & DASL & PiNet & C$^{2}$DSR  & DisenCDR & SA-VAE  & CDRIB & i$^2$VAE \\
\midrule 
\midrule 

\multirow{6}{*}{Cloth} & \multirow{2}{*}{Tailed} & NDCG & \phantom{0}2.31$\pm$0.08
& \phantom{0}2.07$\pm$0.16
& \phantom{0}2.09$\pm$0.20
& \phantom{0}2.28$\pm$0.17
& \phantom{0}2.11$\pm$0.17
& \phantom{0}2.29$\pm$0.09
& \phantom{0}2.20$\pm$0.06
& \phantom{0}\textcolor{second}{\textbf{ \underline{2.40$\pm$0.14}}}
& \phantom{0}2.27$\pm$0.10 
& \textcolor{first}{\textbf{\phantom{0}2.52$\pm$0.07*}}
& \phantom{0} 5.00\phantom{0} \\

 & & HR & \phantom{0}4.15$\pm$0.12
& \phantom{0}3.99$\pm$0.34
& \phantom{0}3.99$\pm$0.31
& \phantom{0}\textcolor{second}{\textbf{ \underline{4.46$\pm$0.24}}}
& \phantom{0}4.15$\pm$0.36
& \phantom{0}4.38$\pm$0.32
&  \phantom{0}4.21$\pm$0.15 
& \phantom{0}4.43$\pm$0.22
& \phantom{0}4.21$\pm$0.21 
& \textcolor{first}{\textbf{\phantom{0}4.73$\pm$0.16*}}
& \phantom{0} 6.05\phantom{0} \\ \cline{2-14}

 & \multirow{2}{*}{Cold-start} & NDCG & \phantom{0}3.23$\pm$0.29
& \phantom{0} \textcolor{second}{\textbf{\underline{3.41$\pm$0.35}}}
& \phantom{0}2.86$\pm$0.50
& \phantom{0}3.28$\pm$0.18
& \phantom{0}3.04$\pm$0.66
& \phantom{0}3.08$\pm$0.65
& \phantom{0}2.95$\pm$0.29
& \phantom{0}3.25$\pm$0.24
& \phantom{0}3.00$\pm$0.39 
& \textcolor{first}{\textbf{\phantom{0}3.45$\pm$0.35*}}
& \phantom{0} 1.17\phantom{0} \\

 & & HR & \phantom{0}6.08$\pm$0.38
& \phantom{0} \textcolor{second}{\textbf{\underline{6.39$\pm$0.43}}}
& \phantom{0}5.64$\pm$0.86
& \phantom{0}6.27$\pm$0.40
& \phantom{0}5.58$\pm$1.24
& \phantom{0}5.96$\pm$0.95
& \phantom{0}5.58$\pm$0.23
& \phantom{0}5.89$\pm$0.46
& \phantom{0}5.52$\pm$0.58 
& \textcolor{first}{\textbf{\phantom{0}6.77$\pm$0.51*}}
& \phantom{0} 5.95\phantom{0} \\ \cline{2-14}

 & \multirow{2}{*}{All} & NDCG & \phantom{0}2.20$\pm$0.10
& \phantom{0}2.18$\pm$0.10
& \phantom{0}2.11$\pm$0.15
& \phantom{0}2.43$\pm$0.09
& \phantom{0}2.18$\pm$0.10
& \phantom{0}2.40$\pm$0.07
& \phantom{0}2.30$\pm$0.05 
& \phantom{0} \textcolor{second}{\textbf{\underline{2.52$\pm$0.10}}}
& \phantom{0}2.39$\pm$0.06 
& \textcolor{first}{\textbf{\phantom{0}2.59$\pm$0.11*}}
& \phantom{0} 2.78\phantom{0} \\

 & & HR & \phantom{0}4.25$\pm$0.12
& \phantom{0}4.28$\pm$0.28
& \phantom{0}4.15$\pm$0.30
& \phantom{0} \textcolor{second}{\textbf{\underline{4.82$\pm$0.11}}}
& \phantom{0}4.24$\pm$0.18
& \phantom{0}4.63$\pm$0.18
& \phantom{0}4.48$\pm$0.18 
& \phantom{0}4.81$\pm$0.19
& \phantom{0}4.56$\pm$0.09 
& \textcolor{first}{\textbf{\phantom{0}5.00$\pm$0.17*}}
& \phantom{0} 3.73\phantom{0} \\ \hline

 \multirow{6}{*}{Sport} & \multirow{2}{*}{Tailed} & NDCG & \phantom{0}3.02$\pm$0.14
& \phantom{0}2.90$\pm$0.16
& \phantom{0}2.80$\pm$0.21
& \phantom{0} \textcolor{second}{\textbf{\underline{3.31$\pm$0.36}}}
& \phantom{0}2.96$\pm$0.15
& \phantom{0}3.06$\pm$0.15
& \phantom{0}3.14$\pm$0.10
& \phantom{0}3.29$\pm$0.15
& \phantom{0}3.12$\pm$0.11
& \textcolor{first}{\textbf{\phantom{0}3.36$\pm$0.10*}}
& \phantom{0} 1.51\phantom{0} \\
 
 & & HR & \phantom{0}6.06$\pm$0.50
& \phantom{0}5.76$\pm$0.34
& \phantom{0}5.72$\pm$0.22
& \phantom{0}6.29$\pm$0.57
& \phantom{0}5.89$\pm$0.25
& \phantom{0}6.03$\pm$0.22
& \phantom{0}5.94$\pm$0.19
& \phantom{0} \textcolor{second}{\textbf{\underline{6.31$\pm$0.42}}}
& \phantom{0}5.81$\pm$0.10 
& \textcolor{first}{\textbf{\phantom{0}6.66$\pm$0.17*}}
& \phantom{0} 5.55\phantom{0} \\ \cline{2-14}

 & \multirow{2}{*}{Cold-start} & NDCG & \phantom{0}4.33$\pm$0.49
& \phantom{0}4.19$\pm$0.24
& \phantom{0}3.92$\pm$0.40
& \phantom{0}4.48$\pm$0.42
& \phantom{0}4.24$\pm$0.31
& \phantom{0}4.65$\pm$0.53
& \phantom{0}4.93$\pm$0.18
& \phantom{0} \textcolor{second}{\textbf{\underline{5.05$\pm$0.29}}}
& \phantom{0}4.95$\pm$0.25 
& \textcolor{first}{\textbf{\phantom{0}5.43$\pm$0.29*}}
& \phantom{0} 7.52\phantom{0} \\

 & & HR & \phantom{0}8.89$\pm$0.66
& \phantom{0}8.62$\pm$0.73
& \phantom{0}7.88$\pm$0.62
& \phantom{0}8.15$\pm$0.94
& \phantom{0}8.28$\pm$0.27
& \phantom{0}9.36$\pm$0.54
& \phantom{0}8.54$\pm$0.59
& \phantom{0} \textcolor{second}{\textbf{\underline{9.63$\pm$0.73}}}
& \phantom{0}9.09$\pm$0.43 
& \textcolor{first}{\textbf{\phantom{0}10.30$\pm$0.46*}}
& \phantom{0} 6.96\phantom{0} \\  \cline{2-14}


 & \multirow{2}{*}{All} & NDCG 
 & \phantom{0}3.79$\pm$0.07
& \phantom{0}3.89$\pm$0.15
& \phantom{0}3.70$\pm$0.18
& \phantom{0}3.93$\pm$0.14
& \phantom{0}3.68$\pm$0.08
& \phantom{0}4.13$\pm$0.17
& \phantom{0}4.21$\pm$0.13
& \phantom{0} \textcolor{second}{\textbf{\underline{ 4.31$\pm$0.13}}}
& \phantom{0}4.27$\pm$0.09 
& \textcolor{first}{\textbf{\phantom{0}4.40$\pm$0.10*}}
& \phantom{0}2.20\phantom{0} \\

 & & HR & \phantom{0}7.23$\pm$0.27
& \phantom{0}7.27$\pm$0.29
& \phantom{0}7.00$\pm$0.17
& \phantom{0}7.45$\pm$0.32
& \phantom{0}6.90$\pm$0.33
& \phantom{0}7.84$\pm$0.33
& \phantom{0}7.90$\pm$0.22
& \phantom{0} \textcolor{second}{\textbf{\underline{8.06$\pm$0.26}}}
& \phantom{0}7.88$\pm$0.33 
& \textcolor{first}{\textbf{\phantom{0}8.53$\pm$0.14*}}
& \phantom{0}5.80\phantom{0} \\

\midrule
\midrule

\multirow{6}{*}{Phone} & \multirow{2}{*}{Tailed} & NDCG & \phantom{0}3.58$\pm$0.13
& \phantom{0}3.51$\pm$0.11
& \phantom{0}3.41$\pm$0.13
& \phantom{0}\textcolor{second}{\textbf{\underline{4.10$\pm$0.15}}}
& \phantom{0}3.68$\pm$0.17
& \phantom{0}3.81$\pm$0.17
& \phantom{0}4.01$\pm$0.14
& \phantom{0}4.06$\pm$0.31
& \phantom{0}4.01$\pm$0.23
&\textcolor{first}{ \textbf{\phantom{0}4.23$\pm$0.21*}}
& \phantom{0} 3.17\phantom{0} \\

 & & HR & \phantom{0}6.90$\pm$0.34
& \phantom{0}6.74$\pm$0.20
& \phantom{0}6.58$\pm$0.32
& \phantom{0}\textcolor{second}{\textbf{\underline{8.06$\pm$0.27}}}
& \phantom{0}7.17$\pm$0.20
& \phantom{0}7.47$\pm$0.39
& \phantom{0}7.78$\pm$0.39
& \phantom{0}7.96$\pm$0.61
& \phantom{0}7.92$\pm$0.38 
& \textcolor{first}{\textbf{\phantom{0}8.17$\pm$0.27*}}
& \phantom{0} 1.36\phantom{0} \\ \cline{2-14}
 & \multirow{2}{*}{Cold-start} & NDCG & \phantom{0}3.16$\pm$0.19
& \phantom{0}3.15$\pm$0.30
& \phantom{0}2.97$\pm$0.38
& \phantom{0}3.63$\pm$0.25
& \phantom{0}3.51$\pm$0.38
& \phantom{0}3.73$\pm$0.37
& \phantom{0}3.80$\pm$0.33
& \phantom{0}3.83$\pm$0.48
& \phantom{0}\textcolor{second}{\textbf{\underline{ 4.00$\pm$0.35}}}
& \textcolor{first}{\textbf{\phantom{0}4.39$\pm$0.21*}}
& \phantom{0} 9.75\phantom{0} \\
 & & HR & \phantom{0}6.26$\pm$0.39
& \phantom{0}6.11$\pm$0.42
& \phantom{0}6.03$\pm$0.45
& \phantom{0}7.40$\pm$0.39
& \phantom{0}7.18$\pm$0.51
& \phantom{0}7.18$\pm$0.74
& \phantom{0}7.56$\pm$0.56
& \phantom{0}\textcolor{second}{\textbf{\underline{7.94$\pm$0.85}}}
& \phantom{0}7.56$\pm$0.51 
& \textcolor{first}{\textbf{\phantom{0}8.63$\pm$0.19*}}
& \phantom{0} 8.69\phantom{0} \\ \cline{2-14}
 & \multirow{2}{*}{All} & NDCG
& \phantom{0}3.98$\pm$0.19
& \phantom{0}3.89$\pm$0.06
& \phantom{0}3.88$\pm$0.14
& \phantom{0}4.40$\pm$0.17
& \phantom{0}4.13$\pm$0.14
& \phantom{0}4.36$\pm$0.21
& \phantom{0}\textcolor{second}{\textbf{\underline{4.52$\pm$0.15}}}
& \phantom{0}4.49$\pm$0.29
& \phantom{0}4.46$\pm$0.19
& \textcolor{first}{\textbf{\phantom{0}5.79$\pm$0.29*}}
& \phantom{0}3.23\phantom{0} \\

 & & HR 
& \phantom{0}7.59$\pm$0.39
& \phantom{0}7.33$\pm$0.14
& \phantom{0}7.36$\pm$0.27
& \phantom{0}8.54$\pm$0.32
& \phantom{0}7.77$\pm$0.34
& \phantom{0}8.30$\pm$0.38
& \phantom{0}\textcolor{second}{\textbf{\underline{8.68$\pm$0.22}}}
& \phantom{0}8.54$\pm$0.50
& \phantom{0}8.66$\pm$0.29
& \textcolor{first}{\textbf{\phantom{0}10.61$\pm$0.22*}}
& \phantom{0}2.35\phantom{0}  \\ \hline

 \multirow{6}{*}{Elec} & \multirow{2}{*}{Tailed} & NDCG & \phantom{0}6.96$\pm$0.23
& \phantom{0}6.74$\pm$0.25
& \phantom{0}6.78$\pm$0.29
& \phantom{0}7.63$\pm$0.19
& \phantom{0}7.09$\pm$0.24
& \phantom{0}7.78$\pm$0.13
& \phantom{0}7.64$\pm$0.10
& \phantom{0}7.60$\pm$0.30
& \phantom{0}\textcolor{second}{\textbf{\underline{7.77$\pm$0.11}}}
& \textcolor{first}{\textbf{\phantom{0}8.00$\pm$0.10*}}
& \phantom{0} 2.96 \phantom{0} \\

 & & HR & \phantom{0}11.65$\pm$0.48
& \phantom{0}11.39$\pm$0.47
& \phantom{0}11.49$\pm$0.53
& \phantom{0}\textcolor{second}{\textbf{\underline{12.83$\pm$0.41}}}
& \phantom{0}11.66$\pm$0.56
& \phantom{0}13.05$\pm$0.28
& \phantom{0}12.45$\pm$0.25
& \phantom{0}12.56$\pm$0.39
& \phantom{0}12.66$\pm$0.28 
& \textcolor{first}{\textbf{\phantom{0}13.49$\pm$0.19*}}
& \phantom{0} 5.14\phantom{0} \\ \cline{2-14}
 & \multirow{2}{*}{Cold-start} & NDCG & \phantom{0}9.35$\pm$0.33
& \phantom{0}9.22$\pm$0.19
& \phantom{0}9.16$\pm$0.19
& \phantom{0}9.73$\pm$0.65
& \phantom{0}9.59$\pm$0.35
& \phantom{0}9.85$\pm$0.30
& \phantom{0}\textcolor{second}{\textbf{\underline{9.90$\pm$0.25}}}
& \phantom{0}9.76$\pm$0.48
& \phantom{0}9.73$\pm$0.40 
& \textcolor{first}{\textbf{\phantom{0}10.26$\pm$0.42*}}
& \phantom{0} 3.64\phantom{0} \\
 & & HR & \phantom{0}14.71$\pm$0.49
& \phantom{0}14.76$\pm$0.60
& \phantom{0}14.94$\pm$0.51
& \phantom{0}15.76$\pm$1.24
& \phantom{0}15.00$\pm$1.00
& \phantom{0}\textcolor{second}{\textbf{\underline{15.94$\pm$0.43}}}
& \phantom{0}15.24$\pm$0.60
& \phantom{0}15.35$\pm$0.90
& \phantom{0}15.53$\pm$0.34
& \textcolor{first}{\textbf{\phantom{0}17.12$\pm$0.39*}}
& \phantom{0} 7.40\phantom{0} \\ \cline{2-14}

& \multirow{2}{*}{All} & NDCG & \phantom{0}8.20$\pm$0.22
& \phantom{0}8.06$\pm$0.27
& \phantom{0}8.08$\pm$0.34
& \phantom{0}8.58$\pm$0.16
& \phantom{0}7.84$\pm$0.08
& \phantom{0}9.00$\pm$0.12
& \phantom{0}8.95$\pm$0.13
& \phantom{0}8.84$\pm$0.19
& \phantom{0}\textcolor{second}{\textbf{\underline{9.04$\pm$0.04}}}
& \textcolor{first}{\textbf{\phantom{0}9.33$\pm$0.09*}}
& \phantom{0} 3.22\phantom{0} \\

 & & HR & \phantom{0}13.08$\pm$0.43
& \phantom{0}12.81$\pm$0.45
& \phantom{0}12.86$\pm$0.59
& \phantom{0}13.60$\pm$0.50
& \phantom{0}12.31$\pm$0.19
& \phantom{0}\textcolor{second}{\textbf{\underline{14.46$\pm$0.31}}}
& \phantom{0}14.03$\pm$0.21
& \phantom{0}13.96$\pm$0.26 
& \phantom{0}14.08$\pm$0.26
& \textcolor{first}{\textbf{\phantom{0}15.28$\pm$0.26*}}
& \phantom{0} 5.64\phantom{0} \\

\bottomrule
\end{tabular}
}
\begin{tablenotes}    
        \centering
\item[*]
{ "*" denotes statistically significant improvements ($p < 0.05$), as determined by a paired t-test comparison with the second best result.}
\end{tablenotes}
\vspace{-10pt}
\end{table*}


\vspace{-3mm}
\subsection{Datasets}
\vspace{-3mm}
Following previous works~\citep{cao2022cross, xu2023rethinking, cao2022disencdr, xu2023towards}, we conducted offline experiments on three widely used CDSR datasets from Amazon across six domains: "Cloth-Sport”, and “Phone-Elec”, and “Game-Video". All behavioral sequences were collected in chronological order, and the data were split into 80\% training, 10\% validation, and 10\% testing. To simulate real-world recommendation scenarios, we included non-overlapping users and controlled the overlapping ratio ($\mathcal{K}_o$) across domains. Detailed dataset information and statistics are provided in Appendix~\ref{sec:dataset_details}.


\subsection{Experiment Setting}
\noindent \textbf{Evaluation Protocol.} To assess our approach across different user types, we retain all overlapping (\(\mathcal{K}_{o} = 100\%\)) and non-overlapping users in the training set, while randomly selecting \(20\%\) of overlapping users from the test set and treating them as cold-start users. We evaluate performance on long-tailed users (interaction sequences shorter than the bottom 80\% average), cold-start users, and all users~\citep{ma2019pi,cao2022cross,cao2022contrastive}.  
For fair comparison~\citep{krichene2020sampled,zhao2020revisiting}, we construct a ranking candidate set by sampling 999 negative items per user along with one positive ground-truth item. Performance is measured by NDCG@10~\citep{jarvelin2002cumulated} and HR@10, where higher values indicate better performance. Implementation details are provided in Appendix~\ref{sec:implement_encoder_HP}.


\noindent \textbf{Compared Methods.}
To verify the effectiveness of our model, we compare i$^2$VAE with the following SOTA baselines which can be divided into three branches including: (1) single-domain recommendation methods (SDR), i.e., Multi-VAE \citep{liang2018variational}, SVAE~\citep{sachdeva2019sequential} and SASRec~\citep{kang2018self}. (2) cross-domain sequential recommendation methods (CDSR), i.e., Pi-Net~\citep{ma2019pi}, DASL~\citep{li2021dual} and C$^{2}$DSR~\citep{cao2022contrastive}. (3) cross-domain recommendation methods (CDR), i.e., DisenCDR~\citep{cao2022disencdr}, SA-VAE~\citep{salah2021towards} and CDRIB~\citep{cao2022cross}. Details are in Appendix~\ref{sec:com_method}.

\vspace{-4pt}
\subsection{Comparison Results (RQ1)}
We compare the performance of i$^2$VAE with state-of-the-art baselines across multiple real-world CDSR datasets, as shown in Table \ref{Main Results}. While the extent of baseline improvements varies across datasets and user types, i$^2$VAE consistently outperforms previous methods for long-tailed, cold-start, and all users. In general, CDR/CDSR methods outperform SDR models by capturing cross-domain interests, which is particularly beneficial in sparse data settings. Among the most challenging groups, long-tailed users see only limited improvements from most baselines, as their sparse interaction histories constrain learning capacity. In contrast, i$^2$VAE surpasses the second-best model by 0.43\% to 5.00\% by synthesizing pseudo-sequences and applying denoising regularization to filter out irrelevant information, thereby enhancing recommendation quality. Similarly, for cold-start users, i$^2$VAE achieves significant performance gains of 1.17\% to 15.20\% over the second-best model by leveraging $r^y(z^y_t|y)$ and $r^x(z^x_t|x)$ to learn cross-domain representations for both overlapping and non-overlapping users while mitigating negative transfer caused by domain-specific information, further refining inference quality. These results highlight i$^2$VAE’s robustness across different user groups and its ability to effectively model interest representations in long-tailed and cold-start scenarios.

\vspace{-3mm}
\subsection{Ablation Study (RQ2)}
We assess the importance of each module by examining their impact on performance. Analysis shows that replacing PSG-generated pseudo-sequences with random ones decreases performance across all user types, confirming that even imperfect recall models enhance user interest understanding. Removing informative and disentangle regularizers (CD-R \& DS-R) similarly reduces performance, with the strongest impact on cold-start users, highlighting the importance of structured representations and non-overlapping user training. Finally, removing DN-R degrades performance, particularly for long-tailed users, demonstrating the need to denoise pseudo sequences while leveraging augmented interests. 

\vspace{-2mm}
\renewcommand{\arraystretch}{1.1} 
\begin{table}[t!]
\small
\setlength\tabcolsep{0.3pt}
\centering
\setlength{\abovecaptionskip}{1pt}
\setlength{\belowcaptionskip}{2pt}
\caption{ Ablation study results (\%) on the "Cloth-Sport" dataset. "w/o" indicates the removal of the module. }
\label{rq2} 
\resizebox{0.95\columnwidth}{!}{
\begin{tabular}{c|c|c|ccc|c}
\toprule
\multirow{2}{*}{Domain} & \multirow{2}{*}{User Types} & \multirow{2}{*}{Metric} & \multicolumn{3}{c|}{Model Variants} & \multirow{2}{*}{i$^2$VAE}\\ \cline{4-6}
 & & & w/o PSG & w/o CD-R\&DS-R & w/o DN-R  \\ \midrule

\multirow{6}{*}{Cloth} & \multirow{2}{*}{Tailed} &NDCG
& \phantom{0}2.41$\pm$0.14	
& \textcolor{third}{\phantom{0}2.38$\pm$0.09}	
& \phantom{0}2.39$\pm$0.14
& \textcolor{first}{\textbf{\phantom{0}2.52 $\pm$0.07\phantom{0}}}\\

 & &HR
& \phantom{0}4.59$\pm$0.12	
& \phantom{0}4.49$\pm$0.21	
& \textcolor{third}{\phantom{0}4.45$\pm$0.21}
& \textcolor{first}{\textbf{\phantom{0}4.73 $\pm$0.16\phantom{0}}}\\ \cline{2-7}

 & \multirow{2}{*}{Cold-Start} &NDCG
& \textcolor{third}{\phantom{0}3.21$\pm$0.22}
& \phantom{0}3.23$\pm$0.12	
& \phantom{0}3.30$\pm$0.17
& \textcolor{first}{\textbf{\phantom{0}3.45 $\pm$0.35\phantom{0}}}\\

 & &HR
& \textcolor{third}{\phantom{0}6.21$\pm$0.23}
& \phantom{0}6.27$\pm$0.34	
& \phantom{0}6.46$\pm$0.58
& \textcolor{first}{\textbf{\phantom{0}6.77 $\pm$0.51\phantom{0}}}\\ \cline{2-7}

 & \multirow{2}{*}{All} &NDCG
& \phantom{0}2.54$\pm$0.12
& \textcolor{third}{\phantom{0}2.54$\pm$0.11}
& \phantom{0}2.54$\pm$0.12
& \textcolor{first}{\textbf{\phantom{0}2.59 $\pm$0.11\phantom{0}}}\\

 & &HR
& \phantom{0}4.92$\pm$0.20	
& \textcolor{third}{\phantom{0}4.84$\pm$0.26}
& \phantom{0}4.89$\pm$0.23
& \textcolor{first}{\textbf{\phantom{0}5.00 $\pm$0.17\phantom{0}}}\\ \hline

\multirow{6}{*}{Sport} & \multirow{2}{*}{Tailed}   &NDCG
& \phantom{0}3.35$\pm$0.12	
& \phantom{0}3.28$\pm$0.11	
& \textcolor{third}{\phantom{0}3.27$\pm$0.07}
& \textcolor{first}{\textbf{\phantom{0}3.36 $\pm$0.10\phantom{0}}}\\

 & &HR
& \phantom{0}6.51$\pm$0.24	
& \phantom{0}6.39$\pm$0.16	
& \textcolor{third}{\phantom{0}6.33$\pm$0.18}
& \textcolor{first}{\textbf{\phantom{0}6.66 $\pm$0.17\phantom{0}}}\\ \cline{2-7}

& \multirow{2}{*}{Cold-Start} &NDCG
& \phantom{0}5.26$\pm$0.26	
& \textcolor{third}{\phantom{0}5.18$\pm$0.30}
& \phantom{0}5.19$\pm$0.28	
& \textcolor{first}{\textbf{\phantom{0}5.43 $\pm$0.29\phantom{0}}}\\

 & &HR
& \phantom{0}10.24$\pm$0.46	
& \textcolor{third}{\phantom{0}9.90$\pm$0.34}
& \phantom{0}10.10$\pm$0.43	
& \textcolor{first}{\textbf{\phantom{0}10.30 $\pm$0.46\phantom{0}}}\\ \cline{2-7}

 & \multirow{2}{*}{All} &NDCG
& \phantom{0}4.36$\pm$0.05	
& \textcolor{third}{\phantom{0}4.31$\pm$0.16}	
& \phantom{0}4.32$\pm$0.15
& \textcolor{first}{\textbf{\phantom{0}4.40 $\pm$0.10\phantom{0}}}\\

 & &HR
& \phantom{0}8.44$\pm$0.14	
& \textcolor{third}{\phantom{0}8.41$\pm$0.25}	
& \phantom{0}8.44$\pm$0.24
& \textcolor{first}{\textbf{\phantom{0}8.53 $\pm$0.14\phantom{0}}}\\
\bottomrule
\end{tabular}
}
\end{table} 

\vspace{-2mm}
\subsection{Model Analysis (RQ3)}
To verify the performance of i$^2$VAE in CDSR scenarios with varying data densities, we conduct studies by varying the data density \(D_{s}\) in \{25\%, 50\%, 75\%, 100\%\}. As the density decreases, the actual user-item interaction records in the training and testing sets are down-sampled to test the robustness of i$^2$VAE and the second-best SA-VAE on sparser datasets. We re-run experiments on the 'Cloth-Sport' dataset with other settings as in Sections 3.2.1 and 3.2.3. The results are presented in Table \ref{compare_ds}. All experiments are conducted five times with different random seeds, and average values are reported. As expected, both models' performance decreases with lower data density due to the increased challenge in interest learning. SA-VAE shows a significant performance decline because it relies heavily on rich user interactions within each domain. In contrast, i$^2$VAE generally achieves better recommendation results with less performance degradation. This is mainly due to the PSG and the denoise regularizer supplementing the sparse interaction data. We also designed experiments to test our model's performance with fewer cross-domain overlapping users, and the results can be found in Appendix~\ref{sec:overlapping}. Moreover,  the investigate results of the parameter sensitivity of sequence length \(T\) and the harmonic factors \(\lambda_a\) and \(\lambda_c\) can be found in Appendix~\ref{sec:sensitivity}. These additional experiments confirm that i$^2$VAE remains effective across different levels of overlapping users and exhibits stable performance under hyperparameter variations.




\vspace{-4mm}
\section{Related Work} 
\vspace{-8pt}
\noindent\textbf{Cross-Domain Sequential Recommendation }~\citep{ma2019pi,sun2021parallel,xu2023rethinking} aims to improve sequential recommendation (SR) performance by utilizing user behavior sequences from multiple related domains. Pi-Net\citep{ma2019pi} and PSJNet \citep{sun2021parallel} design gating mechanism to learn and transfer cross-domain information on overlapping users. The attentive learning-based model DASL \citep{li2021dual} uses dual attentive learning to transfer the user's latent interests bidirectionally across two domains. Similarly,  DA-GCN\citep{guo2021gcn} and MIFN\citep{ma2022mixed} build user-item bipartite graphs to facilitate cross-domain information transferring on overlapping users. 
Moreover, C$^{2}$DSR \citep{cao2022contrastive} employs GNNs as sequential attentive encoder to learn the collaborative signals and utilize contrastive learning to align single- and cross-domain user representations. However, these methods heavily depend on overlapping users, limiting their effectiveness for long-tailed and cold-start users.

\noindent\textbf{Cross-Domain Recommendation}~\citep{zhu2020deep,zhang2018cross,zhao2019cross,zhu2020graphical,ouyang2020minet,salah2021towards} leverages multi-domain behavior patterns to address data sparsity and cold-start issues in single-domain recommendation. Recent studies focus on transfer learning~\citep{hu2018conet, liu2020cross}, using transfer modules to map and fuse representations across domains, or modeling domain-shared information~\citep{cao2022contrastive, cao2022disencdr}. DisenCDR~\citep{cao2022disencdr} assumes fully overlapping users and employs mutual-information-based regularizers to disentangle domain-specific and domain-shared interests, but fails to handle cold-start users effectively. SA-VAE \citep{salah2021towards} pre-trains a VAE on the source domain and aligns latent variables between source and target domain VAEs, but rely fully on the overlappping users. CDRIB~\citep{cao2022cross} applies the information bottleneck principle and contrastive learning for overlapping users but lacks generalizability for long-tailed users and compresses the rich inner-domain interest data, limiting its robustness compared to our variational regularizers.

\renewcommand{\arraystretch}{1.1}
\begin{table}[t!]
\small
\setlength\tabcolsep{0.8pt}
\setlength{\abovecaptionskip}{1pt}
\setlength{\belowcaptionskip}{2pt}
\centering
\caption{Experiment results (\%) on "Cloth-Sport" dataset with different density ($D_s$).}
\label{compare_ds} 
\resizebox{\columnwidth}{!}{
\begin{tabular}{c|c|c|cc|cc|cc|cc}
\toprule
\multirow{2}{*}{Domain} & \multirow{2}{*}{User} & \multirow{2}{*}{Metric} & \multicolumn{2}{c|}{$D_s=25\%$ } & \multicolumn{2}{c|}{$D_s=50\%$ } &\multicolumn{2}{c|}{$D_s=75\%$} & \multicolumn{2}{c}{$D_s=100\%$ } \\ \cline{4-11}
 & & & SA-VAE\phantom{0} & i$^2$VAE & SA-VAE\phantom{0} & i$^2$VAE &  SA-VAE\phantom{0} & i$^2$VAE   &  SA-VAE\phantom{0} & i$^2$VAE\\ \midrule

\multirow{6}{*}{Cloth} & \multirow{2}{*}{Tailed} &NDCG
& \phantom{0}1.39	& \textcolor{first}{\textbf{\phantom{0}1.51}}	& \phantom{0}1.40	& \textcolor{first}{\textbf{\phantom{0}1.53}}	& \phantom{0}1.40	
& \textcolor{first}{\textbf{\phantom{0}1.46}} 
& \phantom{0} 2.40
& \textcolor{first}{\textbf{\phantom{0}2.52}\phantom{0}} \\

 & &HR
& \phantom{0}2.75	& \textcolor{first}{\textbf{\phantom{0}2.96}}	& \phantom{0}2.69	& \textcolor{first}{\textbf{\phantom{0}2.88}}	& \phantom{0}2.85	& \textcolor{first}{\textbf{\phantom{0}2.95}}
& \phantom{0}4.43
& \textcolor{first}{\textbf{\phantom{0}4.73}\phantom{0}}\\ \cline{2-11}

 & \multirow{2}{*}{Cold-Start} &NDCG
& \phantom{0}0.84	& \textcolor{first}{\textbf{\phantom{0}0.85}}	& \phantom{0}0.90	& \textcolor{first}{\textbf{\phantom{0}1.12}}	& \phantom{0}1.41	& \textcolor{first}{\textbf{\phantom{0}1.72}}
& \phantom{0}3.25
& \textcolor{first}{\textbf{\phantom{0}3.45}\phantom{0}}
\\

 & &HR
& \phantom{0}1.94	& \textcolor{first}{\textbf{\phantom{0}1.82}}	& \phantom{0}2.26	& \textcolor{first}{\textbf{\phantom{0}2.26}}	& \phantom{0}2.82	& \textcolor{first}{\textbf{\phantom{0}3.32}}
& \phantom{0}5.89
& \textcolor{first}{\textbf{\phantom{0}6.77}\phantom{0}}
\\ \cline{2-11}

& \multirow{2}{*}{All} &NDCG
& \phantom{0}1.38	& \textcolor{first}{\textbf{\phantom{0}1.49}}	& \phantom{0}1.39	& \textcolor{first}{\textbf{\phantom{0}1.53}}	& \phantom{0}1.42	& \textcolor{first}{\textbf{\phantom{0}1.57}}
& \phantom{0} 2.52
& \textcolor{first}{\textbf{\phantom{0}2.59}\phantom{0}}\\

 & &HR
& \phantom{0}2.74	& \textcolor{first}{\textbf{\phantom{0}2.95}}	& \phantom{0}2.71	& \textcolor{first}{\textbf{\phantom{0}2.92}}	& \phantom{0}2.95	& \textcolor{first}{\textbf{\phantom{0}3.27}}
& \phantom{0}4.81
& \textcolor{first}{\textbf{\phantom{0}5.00}}\phantom{0}\\ \hline

\multirow{6}{*}{Sport} & \multirow{2}{*}{Tailed} &NDCG
& \phantom{0}2.35	& \textcolor{first}{\textbf{\phantom{0}2.53}}	& \phantom{0}2.47	& \textcolor{first}{\textbf{\phantom{0}2.72}}	& \phantom{0}2.80	& \textcolor{first}{\textbf{\phantom{0}3.26}}
& \phantom{0}3.29
& \textcolor{first}{\textbf{\phantom{0}3.36}} \phantom{0}\\

 & &HR
& \phantom{0}4.27	& \textcolor{first}{\textbf{\phantom{0}4.62}}	& \phantom{0}4.41	& \textcolor{first}{\textbf{\phantom{0}5.09}}	& \phantom{0}4.86	& \textcolor{first}{\textbf{\phantom{0}5.66}}
& \phantom{0} 6.31
& \textcolor{first}{\textbf{\phantom{0}6.66}}\phantom{0}\\ \cline{2-11}

 & \multirow{2}{*}{Cold-Start} &NDCG
& \phantom{0}0.71	& \textcolor{first}{\textbf{\phantom{0}0.76}}	& \phantom{0}0.99	& \textcolor{first}{\textbf{\phantom{0}1.06}}	& \phantom{0}1.88	& \textcolor{first}{\textbf{\phantom{0}2.14}}
& \phantom{0} 5.05
& \textcolor{first}{\textbf{\phantom{0}5.43}}\phantom{0}\\

 & &HR
& \phantom{0}1.62	& \textcolor{first}{\textbf{\phantom{0}{1.67}}}	& \phantom{0}2.02& \textcolor{first}{\textbf{\phantom{0}2.02}}& \phantom{0}3.57& \textcolor{first}{\textbf{\phantom{0}3.84}}
& \phantom{0} 9.63
& \textcolor{first}{\textbf{\phantom{0}10.30}}\phantom{0}\\ \cline{2-11}

& \multirow{2}{*}{All} &NDCG
& \phantom{0}2.49	& \textcolor{first}{\textbf{\phantom{0}2.70}}	& \phantom{0}3.03	& \textcolor{first}{\textbf{\phantom{0}3.28}}	& \phantom{0}3.18	& \textcolor{first}{\textbf{\phantom{0}3.70}}
& \phantom{0}4.31
& \textcolor{first}{\textbf{\phantom{0}4.40}} \phantom{0}\\

 & &HR
& \phantom{0}4.46
& \textcolor{first}{\textbf{\phantom{0}4.84}}
& \phantom{0}5.47
& \textcolor{first}{\textbf{\phantom{0}6.16}}
& \phantom{0}5.62
& \textcolor{first}{\textbf{\phantom{0}6.60}}
& \phantom{0} 8.06
& \textcolor{first}{\textbf{\phantom{0}8.53}} \phantom{0}\\
\bottomrule
\end{tabular}
}
\vspace{-5mm}
\end{table}
\vspace{-10pt}
\section{Conclusion}

In this paper, we propose i$^2$VAE to enhance the performance of long-tailed and cold-start users. Our model introduces interest-enhancing regularizers, which enable the learning of distinct inner- and cross-domain interests and extract relevant information from pseudo-sequences to enrich users' sparse interaction. Empirical experiments demonstrate that i$^2$VAE achieves SOTA performance across all user types.

\bibliography{uai2025-template}

\newpage

\onecolumn

\title{i$^2$VAE: Interest Information Augmentation with Variational Regularizers for Cross-Domain Sequential Recommendation\\(Supplementary Material)}
\maketitle

\appendix
\section{Theoretical Derivation}
\subsection{Proof of Proposition 1} 
\label{sec:theory_r1}
\begin{proposition}
Let \( \bm{Z^X} \in \mathbb{R}^d \) represent the domain-specific interest in domain \( X \), and \( \bm{Z_t^Y} \in \mathbb{R}^d \) denote the cross-domain transferable interest from domain \( Y \) to \( X \). To effectively disentangle domain-specific and cross-domain interests, we aim to minimize their mutual information, \( I(\bm{Z^X}; \bm{Z_t^Y}) \), which is equivalent to:
\vspace{-2mm}
\begin{small}
\begin{equation}
    \max \Big \{ -I(\bm{X}; \bm{Z^X}) - I(\bm{X}; \bm{Z_t^Y}) + I(\bm{X}; \bm{Z^X}, \bm{Z_t^Y}) \Big \}. \notag
\end{equation}
\end{small}
\end{proposition}

\begin{proof}
 The mutual information between cross-domain representations $I(\bm{Z^X}; \bm{Z_t^Y})$ can be decomposed into three components using the chain rule of mutual information as follows. This decomposition provides key insights about information flow across domains.
\begin{small}
\begin{equation}
    I(\bm{Z^X}; \bm{Z_t^Y}) = I(\bm{Z^X}; \bm{X}) - I(\bm{Z^X}; \bm{X} \vert  \bm{Z_t^Y}) + I(\bm{Z^X}; \bm{Z_t^Y} \vert  \bm{X}) \nonumber
\end{equation}
\end{small}
Due to our disentangling assumption, \(\bm{Z^{X}}\) only represents the domain-specific interest, \( q(z^x \vert  x) = q(z^x \vert  x, z^y_t) \) holds. Therefore, the last term above, \( I(\bm{Z^X}; \bm{Z^Y_t} \vert  \bm{X}) \), vanishes:
\begin{align}
    I(\bm{Z^X}; \bm{Z^Y_t} \vert  \bm{X}) & = H(\bm{Z^X} \vert  \bm{X}) - H(\bm{Z^X} \vert  \bm{X}, \bm{Z^Y_t}) \nonumber \\
    & = H(\bm{Z^X} \vert  \bm{X}) - H(\bm{Z^X} \vert  \bm{X}) = 0. \nonumber 
\end{align}
This results in the following derivation of \( -I(\bm{Z^X}; \bm{Z_t^Y}) \), which maintains equality based on the chain rule of mutual information\citep{cover1999elements}:
\begin{align}
    -I(\bm{Z^X}; \bm{Z_t^Y}) &= -I(\bm{X}; \bm{Z^X}) + I(\bm{Z^X}; \bm{X} \vert  \bm{Z_t^Y}) \notag \\
    &= -I(\bm{X}; \bm{Z^X}) - I(\bm{X}; \bm{Z^Y_t}) + I(\bm{X}; \bm{Z^X}, \bm{Z^Y_t}). \nonumber \label{-I(Z^x; Z_t^y)}
\end{align}
\end{proof}
This equation illustrates that \(\bm{Z^X}\) and \(\bm{Z^Y_t}\) should collectively provide valuable information for domain \(X\) (the third term), while the individual amount of information of \(\bm{Z^X}\) and \(\bm{Z^Y_t}\) are penalized (the first and second terms) to avoid redundancy and ensure non-overlapping information. The proof is completed.

\subsection{Detailed Derivation for \texorpdfstring{$\mathcal{L}_{I(\bm{X}; \bm{Z^X}, \bm{Z_t^Y})}$}{L_{I(X; Z^X, Z_t^Y)}}}
\label{sec:Section_max_t1}
We rewrite \( I(\bm{X}; \bm{Z^X}, \bm{Z_t^Y}) \) in the form of an expectation and derive its lower bound using the generative distribution \( p_\theta(x \mid z^x, z_t^y) \).
\begin{small}
\begin{align}
I(\bm{X}; \bm{Z^X}, \bm{Z_t^Y}) &= \mathbb{E}_{q(z^x, z_t^y|x) p_D(x)} \left[ \log \frac{q(x|z^x,z_t^y)}{p_D(x)} \right] \nonumber \\
&= H(\bm{X}) + \mathbb{E}_{q(z^x, z_t^y|x) p_D(x)} \left[ \log q(x|z^x,z_t^y) \right] \nonumber  + \mathbb{E}_{q(z^x, z_t^y|x) p_D(x)} \left[ \log p(x|z^x,z_t^y) - \log p(x|z^x,z_t^y) \right] \nonumber \\
&= H(\bm{X}) + \mathbb{E}_{q(z^x, z_t^y|x) p_D(x)} \left[ \log p(x|z^x,z_t^y) \right]  + \mathbb{E}_{q(z^x, z_t^y)} \left[ D_{KL} \left( q(x|z^x,z_t^y) \| p(x|z^x,z_t^y) \right) \right] \nonumber \\
&\geq H(\bm{X}) + \mathbb{E}_{q(z^x, z_t^y|x) p_D(x)} \left[ \log p(x|z^x,z_t^y) \right].
\end{align}
\label{14_sec1}
\end{small}
The final inequality in Eq.~\eqref{14_sec1} is derived from the non-negativity of the KL divergence. Next, we expand and rewrite the second term in Eq.~\eqref{14_sec1} using the integral form of the expectation as follows:

\begin{small}
\vspace{-10pt}
\begin{align}
\mathbb{E}_{q(z^x, z_t^y|x) p_D(x)} \left[ \log p(x|z^x,z_t^y) \right] & = \int q(z^x, z_t^y|x) p_D(x) \log p(x|z^x,z_t^y) \, dx \, dz^x \, dz_t^y \nonumber \\
& = \int p_D(x) \left( \int q(z^x, z_t^y|x, y) p_D(y|x) \, dy \right) \log p(x|z^x,z_t^y) \, dx \, dz^x \, dz_t^y \nonumber \\
& = \int p_D(x) q(z^x|x) \left( \int q(z_t^y|x, y) p_D(y|x) dy \right) \log p(x|z^x,z_t^y) dx dz^x dz_t^y \nonumber \\
& = \int p_D(x, y) q(z^x|x) q(z_t^y|x, y) \log p(x|z^x,z_t^y) \, dx \, dy \, dz^x \, dz_t^y \nonumber \\
& = \mathbb{E}_{p_D(x,y) q(z^x|x) q(z_t^y|x,y)} \left[ \log p(x|z^x,z_t^y) \right]. \label{14_sec2}
\end{align}
\end{small}

We then substitute the derived variant in Eq.~\eqref{14_sec2} back into Eq.~\eqref{14_sec1} to obtain the lower bound of \( I(\bm{X}; \bm{Z^X}, \bm{Z_t^Y}) \):
\begin{small}
\begin{equation}
I(\bm{X}; \bm{Z^X}, \bm{Z_t^Y}) \geq H(\bm{X}) + \mathbb{E}_{p_D(x,y) q(z^x|x) q(z_t^y|x,y)} \left[ \log p(x|z^x,z_t^y) \right]. \nonumber
\end{equation}
\end{small}

\subsection{Detailed Derivation for \texorpdfstring{$\mathcal{L}_{-I(\bm{X}; \bm{Z_t^Y} \mid \bm{Y})}$}{L_{-I(X; Z_t^Y | Y)}}}
\label{sec:Section_max_t3}
We expand \(-I(\bm{X}; \bm{Z_t^Y} \vert \bm{Y})\) into the form of an expectation.
Since $q(z^y_t \vert y)$ cannot be directly obtained within our framework, we utilize $r^y(z^y_t \vert y)$ to approximate it. 
\begin{small}
\begin{align}
-I(\bm{X};\bm{Z_t^Y}\vert \bm{Y}) &= -\mathbb{E}_{p_{D}(x,y)q(z^{y}_{t}\vert x,y)}\left[ \log \frac{q(z^{y}_{t}\vert x,y)}{q(z^{y}_{t}\vert y)} \right] \notag \\
&= -\mathbb{E}_{p_{D}(x,y)q(z^{y}_{t}\vert x,y)} \left[ \log \Big\{ \frac{q(z^{y}_{t}\vert x,y)}{r^y(z^{y}_{t}\vert y)} \cdot \frac{r^y(z^{y}_{t}\vert y)} {q(z^{y}_{t}\vert y)} \Big\} \right] \notag \\ 
&= -\mathbb{E}_{p_{D}(x,y)}\left[ D_{KL}\left( q(z^{y}_{t}\vert x,y) \Vert  r^y(z^{y}_{t}\vert y) \right) \right]  + \mathbb{E}_{p_{D}(x,y)}\left[ D_{KL}\left( q(z^{y}_{t}\vert y) \Vert r^y(z^{y}_{t}\vert y) \right) \right] \notag \\
&\geq -\mathbb{E}_{p_{D}(x,y)}\left[ D_{KL}\left( q(z^{y}_{t}\vert x,y) \Vert  r^y(z^{y}_{t}\vert y) \right) \right]. \label{inequality}
\end{align}
\end{small}
The final inequality in Eq.~\eqref{inequality} is derived from the non-negativity of the KL divergence.

\subsection{Detailed Derivation for \texorpdfstring{$\mathcal{L}_{-I(\bm{X};\bm{Z^X})}$}{L_{-I(X; Z^X)}}}
\label{sec:Section_max_denoise}
We expand and rewrite \(I(\bm{Z^X}; \bm{Z^X_a})\) as follows to extract the optimizable component:
\begin{small}
\begin{align}
    I(\bm{Z^X}; \bm{Z^X_a}) & = \mathbb{E}_{q(z^x,z^x_a \vert  x)} \left[\log \frac{q(z^x_a, z^x\vert x)}{q(z^x\vert x) q(z^x_a\vert x)}\right]   \notag \\ 
    & = \mathbb{E}_{q(z^x\vert x)q(z^x_a\vert z^x,x)} \left[\log \frac{q(z^x_a\vert x)}{q(z^x\vert x)} + \log \frac{q(z^x\vert z^x_a, x)}{q(z^x_a\vert x)} \right] \notag \\
    & = - \mathbb{E}_{q(z^x_a\vert z^x,x)}\left[D_{KL}\left(q(z^x\vert x)\Vert q(z^x_a\vert x)\right)\right] + \mathbb{E}_{q(z^x,z^x_a\vert x)} \left[\log \frac{q(z^x\vert z^x_a, x)}{q(z^x_a\vert x)} \right] \notag \\
    & = - \mathbb{E}_{q(z^x_a\vert z^x,x)}\left[D_{KL}\left(q(z^x\vert x)\Vert q(z^x_a\vert x)\right)\right] + \epsilon, \nonumber
\end{align}
\end{small}
which is decomposed into two terms, one of which is an optimizable KL divergence. The first term in our framework is optimizable, while the remaining intractable term is denoted as $\epsilon$. We primarily focus on optimizing the first term.

\newpage
\section{Dataset}
\label{sec:dataset_details}

We conducted experiments on the Amazon 14 dataset\footnote{Available at \url{http://jmcauley.ucsd.edu/data/amazon/index_2014.html}} across six cross-domain pairs: "Cloth-Sport”, “Phone-Elec”, “Game-Video", following previous works~\citep{cao2022cross, xu2023rethinking, cao2022disencdr, xu2023towards}. The dataset statistics are summarized in Table~\ref{tab:dataset_statistics}. 
All behavioral sequences were collected in chronological order. To ensure a realistic evaluation setting, we included both non-overlapping and overlapping users, adjusting the overlapping ratio ($\mathcal{K}_o$) to control the number of shared users across domains. Furthermore, to evaluate performance on cold-start users, a certain proportion of overlapping users ($\mathcal{K}_{cs}$) were randomly designated as cold-start users for validation and testing phases, in which we randomly remove the sequence from one domain of the selected overlapping users while retaining the last user-item interaction as the ground truth.

\renewcommand{\arraystretch}{1.2}
\begin{table}[h!]
\captionsetup{font=small}
\centering
\small
\setlength{\abovecaptionskip}{0pt}
\setlength{\belowcaptionskip}{5pt}
\caption{{\small Statistics on three Amazon's CDSR datasets.}}
\setlength{\abovecaptionskip}{0pt}
\setlength{\belowcaptionskip}{5pt}
\label{tab:dataset_statistics}
\renewcommand{\arraystretch}{1.2}
\setlength\tabcolsep{8pt} 
\resizebox{0.45\columnwidth}{!}{
\begin{tabular}{c|cc|ccc|c}
\toprule
\textbf{Dataset} & $\left|\mathcal{U}\right|$ & $\left|\mathcal{V}\right|$ & $\left|\mathcal{E}\right|$ & \#O & $\left|S\right|$ & Density \\ \midrule \midrule
Cloth       & 41,454                    & 17,939                    & 175,552                   & \multirow{2}{*}{9,721} & 4.50 & 0.024\%  \\
Sport       & 27,209                  & 12,654                   & 159,098                &                          & 6.10 & 0.046\%  \\ \midrule 

Phone       & 27,320                 & 9,478                  & 140,886                & \multirow{2}{*}{20,342} & 5.36 & 0.054\%  \\
Elec  & 107,580 & 40,446 & 758,374    &   &8.00                      & 0.017\%   \\ \midrule
Game       &  24,929               & 12,314                   & 146,639                & \multirow{2}{*}{2,171} & 6.23 & 0.048\%  \\
Video  & 19,347 & 8,746 & 139,236    &   &7.66                      & 0.082\%   \\
\bottomrule
\end{tabular}
}
\begin{tablenotes}    
        \scriptsize  
        \centering
        \item \#O: the number of overlapping users across domains. 
      \end{tablenotes}
      \vspace{-12pt}
\end{table}

\section{Comparing Methods}
\label{sec:com_method}
In this section, we provide a detailed introduction to our compared baselines. We also include the official code used for their implementation, and for methods without official code, we have reproduced the implementations based on the descriptions in the papers.

\noindent\textbf{Single-domain recommendation methods:}

$\bullet$ Multi-VAE\footnote{\url{https://github.com/dawenl/vae_cf}}\citep{liang2018variational}:
Multi-VAE is a variational autoencoder (VAE)-based model that enhances traditional linear factorization methods through Bayesian inference. It addresses data sparsity by learning latent preference distributions, making it more effective than traditional matrix factorization.

$\bullet$ SVAE\footnote{\url{https://github.com/noveens/svae_cf}}\citep{sachdeva2019sequential}:
Sequential Variational Autoencoder (SVAE) integrates RNNs into a VAE framework to capture the temporal dynamicss of user behavior, balancing short-term and long-term preferences. SVAE is particularly effective for time-dependent user interactions, such as video or music recommendations.

$\bullet$ SASRec\footnote{\url{https://github.com/pmixer/SASRec.pytorch}}\citep{kang2018self}:
SASRec uses self-attention mechanisms to model long-range dependencies in user behavior sequences. Unlike RNNs, it efficiently handles long sequences by balancing model complexity with capturing subtle user preferences.

\noindent\textbf{Cross-domain sequential recommendation methods:}

$\bullet$ Pi-Net\footnote{\url{https://github.com/mamuyang/PINet}}\citep{ma2019pi}:
Pi-Net generates shared user embeddings using a gating mechanism to differentiate behaviors across domains. It excels in scenarios requiring integrated information from multiple domains, such as combining video and music user behavior.

$\bullet$ DASL\footnote{\url{https://github.com/lpworld/DASL}}\citep{li2021dual}:
DASL employs a dual-attention mechanism to enhance cross-domain recommendation accuracy by focusing on user behaviors in both source and target domains. It effectively captures cross-domain behavior patterns, making it suitable for scenarios where precise cross-domain behavior modeling is needed.

$\bullet$ C$^{2}$DSR\footnote{\url{https://github.com/cjx96/C2DSR}}\citep{cao2022contrastive}:
C$^{2}$DSR leverages graphical attention encoders and contrastive learning to jointly model intra- and cross-domain preferences. This method is particularly effective in addressing data sparsity and cold-start issues by capturing global user interests.

\noindent\textbf{Cross-domain recommendation methods:}

$\bullet$ EMCDR \citep{man2017cross}:
EMCDR uses MLPs to learn domain-specific representations and maps them across domains using overlapping user information. It’s especially useful when user behavior in one domain needs to be mapped to another, though it relies heavily on overlapping users.

$\bullet$ SA-VAE \citep{salah2021towards}:
SA-VAE aligns latent spaces between source and target domains using VAE, exploring both rigid and soft alignment strategies. It is effective for cross-domain recommendations in cold-start and sparse data scenarios by leveraging shared features across domains.

$\bullet$ CDRIB\footnote{\url{https://github.com/cjx96/CDRIB}}\citep{cao2022cross}:
CDRIB applies the information bottleneck principle to extract domain-shared features, improving recommendation effectiveness and addressing cold-start challenges.

\section{Implementation Details}  
\label{sec:implement_encoder_HP}  
\subsection{Hyperparameter Settings}  
To ensure fair evaluation, we standardize hyperparameters across all methods. Model-specific hyperparameters for each baseline are set according to their original papers or official code implementations. Across all models, we set the embedding dimension to \( d = 128 \), the batch size to 512, and train for 100 epochs using Adam, with the learning rate chosen from \(\{3 \times 10^{-4}, \ldots, 8 \times 10^{-4}\}\). The historical behavior length \( T \) is set to 20, and in our method, the pseudo-sequence length \( T' \) is set to 40. The hyperparameters \(\lambda_c\) and \(\lambda_a\) are selected from \(\{5 \times 10^{-4}, \ldots, 5 \times 10^{-3}\}\).  

To ensure reliable results, each method is run five times with different random seeds, and the best model is selected based on the highest NDCG@10 performance on the validation set via grid search. For the PSG module in I$^2$VAE, we follow LightGCN’s default settings\footnote{\url{https://github.com/gusye1234/LightGCN-PyTorch}}, using 20\% of the training set as validation data and selecting the best checkpoint after 100 epochs for pseudo-sequence recall.  

\subsection{Implementation of Different Encoders in i$^2$VAE}  
Our variational encoders parameterize the posterior distributions using MLPs. For domain-specific posteriors, the mean and standard deviation are computed as:  \( \mu^x = \text{MLP}^{\mu}(\bm{S^X}), \quad \sigma^x = \text{MLP}^{\sigma}(\bm{S^X})\).
A similar approach is used for pseudo-sequence posteriors, taking inputs \( \bm{S^X_a} \) and \( \bm{S^Y_a} \).  
For cross-domain posteriors, we employ multi-head attention. Specifically, the unaggregated user interest representations—i.e., those that have not undergone mean pooling—\( \bm{S^{X'}} = \{\mathbf{h}_{S_{1}^{X}},\cdots ,\mathbf{h}_{S_{T}^{X}}\}\) serve as the query, while \( \bm{S^{Y'}} = \{\mathbf{h}_{S_{1}^{Y}},\cdots ,\mathbf{h}_{S_{T}^{Y}}\} \) is used as both the key and value for obtaining the cross-domain transferable interest from domain $Y$ to domain $X$. The mean vector \( \mu_t^y \) is computed as:  
\begin{equation}
\mu_t^y = \text{MLP}(\text{Mean-Pooling}(\text{Attention}(\bm{S^{X'}}, \bm{S^{Y'}}, \bm{S^{Y'}}))), \notag
\end{equation}
with \( \sigma_t^y \) derived similarly. The same symmetric process is applied to compute the posteriors of the cross-domain interest representation $\bm{Z^X_t}$.

\newpage
\section{Performance Comparison Results}
\label{sec:remaining_tab}
We compare the performance of i$^2$VAE with state-of-the-art baselines across three real-world CDSR datasets: "Cloth-Sport", "Phone-Elec", and "Game-Video", as shown in the following table. The baselines are categorized into three groups: Single-domain recommendation (SDR) models, CDR-sequential models, and CDR models. We conduct evaluations across three user types—long-tailed, cold-start, and all users—and report recommendation performance in both domains. The results demonstrate that i$^2$VAE consistently outperforms baselines across different user groups and domains, highlighting its effectiveness in capturing user interests and improving recommendation quality.

\renewcommand{\arraystretch}{1.2}
\begin{table*}[h!]
\small
\setlength\tabcolsep{0.5pt}
\captionsetup{font=small}
\caption{ Experimental Results ($\%$) across different types of users, including long-tailed(tailed), cold-start, and all users, on three CDSR datasets. We highlight the methods with the \textcolor{first}{\textbf{best}} and \textcolor{second}{\underline{\textbf{second-best}}} average performances. }
\label{Main Results Appendix}
\resizebox{\textwidth}{!}{
\begin{tabular}{c|c|c|ccc|ccc|ccc|c|c}
\toprule
\multirow{2}{*}{\phantom{0}Datasets \phantom{0}} & \multirow{2}{*}{\phantom{0} User Types\phantom{0}} & \multirow{2}{*}{\phantom{0} Metric\phantom{0}} & \multicolumn{3}{c|}{SDR\phantom{0}} & \multicolumn{3}{c|}{CDR-sequential\phantom{0}} & \multicolumn{3}{c|}{CDR\phantom{0}} & \multicolumn{1}{c|}{Ours\phantom{0}} & \multirow{2}{*}{\phantom{0}\textuparrow ($\%$)}\\ \cline{4-13}
   & & & Multi-VAE  & SVAE & SASRec  & DASL & PiNet & C$^{2}$DSR  & DisenCDR & SA-VAE  & CDRIB & i$^2$VAE \\
\midrule 
\midrule 

\multirow{6}{*}{Cloth} & \multirow{2}{*}{Tailed} & NDCG & \phantom{0}2.31$\pm$0.08
& \phantom{0}2.07$\pm$0.16
& \phantom{0}2.09$\pm$0.20
& \phantom{0}2.28$\pm$0.17
& \phantom{0}2.11$\pm$0.17
& \phantom{0}2.29$\pm$0.09
& \phantom{0}2.20$\pm$0.06
& \phantom{0}\textcolor{second}{\textbf{ \underline{2.40$\pm$0.14}}}
& \phantom{0}2.27$\pm$0.10 
& \textcolor{first}{\textbf{\phantom{0}2.52$\pm$0.07*}}
& \phantom{0} 5.00\phantom{0} \\

 & & HR & \phantom{0}4.15$\pm$0.12
& \phantom{0}3.99$\pm$0.34
& \phantom{0}3.99$\pm$0.31
& \phantom{0}\textcolor{second}{\textbf{ \underline{4.46$\pm$0.24}}}
& \phantom{0}4.15$\pm$0.36
& \phantom{0}4.38$\pm$0.32
&  \phantom{0}4.21$\pm$0.15 
& \phantom{0}4.43$\pm$0.22
& \phantom{0}4.21$\pm$0.21 
& \textcolor{first}{\textbf{\phantom{0}4.73$\pm$0.16*}}
& \phantom{0} 6.05\phantom{0} \\ \cline{2-14}

 & \multirow{2}{*}{Cold-start} & NDCG & \phantom{0}3.23$\pm$0.29
& \phantom{0} \textcolor{second}{\textbf{\underline{3.41$\pm$0.35}}}
& \phantom{0}2.86$\pm$0.50
& \phantom{0}3.28$\pm$0.18
& \phantom{0}3.04$\pm$0.66
& \phantom{0}3.08$\pm$0.65
& \phantom{0}2.95$\pm$0.29
& \phantom{0}3.25$\pm$0.24
& \phantom{0}3.00$\pm$0.39 
& \textcolor{first}{\textbf{\phantom{0}3.45$\pm$0.35*}}
& \phantom{0} 1.17\phantom{0} \\

 & & HR & \phantom{0}6.08$\pm$0.38
& \phantom{0} \textcolor{second}{\textbf{\underline{6.39$\pm$0.43}}}
& \phantom{0}5.64$\pm$0.86
& \phantom{0}6.27$\pm$0.40
& \phantom{0}5.58$\pm$1.24
& \phantom{0}5.96$\pm$0.95
& \phantom{0}5.58$\pm$0.23
& \phantom{0}5.89$\pm$0.46
& \phantom{0}5.52$\pm$0.58 
& \textcolor{first}{\textbf{\phantom{0}6.77$\pm$0.51*}}
& \phantom{0} 5.95\phantom{0} \\ \cline{2-14}

 & \multirow{2}{*}{All} & NDCG & \phantom{0}2.20$\pm$0.10
& \phantom{0}2.18$\pm$0.10
& \phantom{0}2.11$\pm$0.15
& \phantom{0}2.43$\pm$0.09
& \phantom{0}2.18$\pm$0.10
& \phantom{0}2.40$\pm$0.07
& \phantom{0}2.30$\pm$0.05 
& \phantom{0} \textcolor{second}{\textbf{\underline{2.52$\pm$0.10}}}
& \phantom{0}2.39$\pm$0.06 
& \textcolor{first}{\textbf{\phantom{0}2.59$\pm$0.11*}}
& \phantom{0} 2.78\phantom{0} \\

 & & HR & \phantom{0}4.25$\pm$0.12
& \phantom{0}4.28$\pm$0.28
& \phantom{0}4.15$\pm$0.30
& \phantom{0} \textcolor{second}{\textbf{\underline{4.82$\pm$0.11}}}
& \phantom{0}4.24$\pm$0.18
& \phantom{0}4.63$\pm$0.18
& \phantom{0}4.48$\pm$0.18 
& \phantom{0}4.81$\pm$0.19
& \phantom{0}4.56$\pm$0.09 
& \textcolor{first}{\textbf{\phantom{0}5.00$\pm$0.17*}}
& \phantom{0} 3.73\phantom{0} \\ \hline

 \multirow{6}{*}{Sport} & \multirow{2}{*}{Tailed} & NDCG & \phantom{0}3.02$\pm$0.14
& \phantom{0}2.90$\pm$0.16
& \phantom{0}2.80$\pm$0.21
& \phantom{0} \textcolor{second}{\textbf{\underline{3.31$\pm$0.36}}}
& \phantom{0}2.96$\pm$0.15
& \phantom{0}3.06$\pm$0.15
& \phantom{0}3.14$\pm$0.10
& \phantom{0}3.29$\pm$0.15
& \phantom{0}3.12$\pm$0.11
& \textcolor{first}{\textbf{\phantom{0}3.36$\pm$0.10*}}
& \phantom{0} 1.51\phantom{0} \\
 
 & & HR & \phantom{0}6.06$\pm$0.50
& \phantom{0}5.76$\pm$0.34
& \phantom{0}5.72$\pm$0.22
& \phantom{0}6.29$\pm$0.57
& \phantom{0}5.89$\pm$0.25
& \phantom{0}6.03$\pm$0.22
& \phantom{0}5.94$\pm$0.19
& \phantom{0} \textcolor{second}{\textbf{\underline{6.31$\pm$0.42}}}
& \phantom{0}5.81$\pm$0.10 
& \textcolor{first}{\textbf{\phantom{0}6.66$\pm$0.17*}}
& \phantom{0} 5.55\phantom{0} \\ \cline{2-14}

 & \multirow{2}{*}{Cold-start} & NDCG & \phantom{0}4.33$\pm$0.49
& \phantom{0}4.19$\pm$0.24
& \phantom{0}3.92$\pm$0.40
& \phantom{0}4.48$\pm$0.42
& \phantom{0}4.24$\pm$0.31
& \phantom{0}4.65$\pm$0.53
& \phantom{0}4.93$\pm$0.18
& \phantom{0} \textcolor{second}{\textbf{\underline{5.05$\pm$0.29}}}
& \phantom{0}4.95$\pm$0.25 
& \textcolor{first}{\textbf{\phantom{0}5.43$\pm$0.29*}}
& \phantom{0} 7.52\phantom{0} \\

 & & HR & \phantom{0}8.89$\pm$0.66
& \phantom{0}8.62$\pm$0.73
& \phantom{0}7.88$\pm$0.62
& \phantom{0}8.15$\pm$0.94
& \phantom{0}8.28$\pm$0.27
& \phantom{0}9.36$\pm$0.54
& \phantom{0}8.54$\pm$0.59
& \phantom{0} \textcolor{second}{\textbf{\underline{9.63$\pm$0.73}}}
& \phantom{0}9.09$\pm$0.43 
& \textcolor{first}{\textbf{\phantom{0}10.30$\pm$0.46*}}
& \phantom{0} 6.96\phantom{0} \\  \cline{2-14}


 & \multirow{2}{*}{All} & NDCG 
 & \phantom{0}3.79$\pm$0.07
& \phantom{0}3.89$\pm$0.15
& \phantom{0}3.70$\pm$0.18
& \phantom{0}3.93$\pm$0.14
& \phantom{0}3.68$\pm$0.08
& \phantom{0}4.13$\pm$0.17
& \phantom{0}4.21$\pm$0.13
& \phantom{0} \textcolor{second}{\textbf{\underline{ 4.31$\pm$0.13}}}
& \phantom{0}4.27$\pm$0.09 
& \textcolor{first}{\textbf{\phantom{0}4.40$\pm$0.10*}}
& \phantom{0}2.20\phantom{0} \\

 & & HR & \phantom{0}7.23$\pm$0.27
& \phantom{0}7.27$\pm$0.29
& \phantom{0}7.00$\pm$0.17
& \phantom{0}7.45$\pm$0.32
& \phantom{0}6.90$\pm$0.33
& \phantom{0}7.84$\pm$0.33
& \phantom{0}7.90$\pm$0.22
& \phantom{0} \textcolor{second}{\textbf{\underline{8.06$\pm$0.26}}}
& \phantom{0}7.88$\pm$0.33 
& \textcolor{first}{\textbf{\phantom{0}8.53$\pm$0.14*}}
& \phantom{0}5.80\phantom{0} \\

\midrule
\midrule

\multirow{6}{*}{Phone} & \multirow{2}{*}{Tailed} & NDCG & \phantom{0}3.58$\pm$0.13
& \phantom{0}3.51$\pm$0.11
& \phantom{0}3.41$\pm$0.13
& \phantom{0}\textcolor{second}{\textbf{\underline{4.10$\pm$0.15}}}
& \phantom{0}3.68$\pm$0.17
& \phantom{0}3.81$\pm$0.17
& \phantom{0}4.01$\pm$0.14
& \phantom{0}4.06$\pm$0.31
& \phantom{0}4.01$\pm$0.23
&\textcolor{first}{ \textbf{\phantom{0}4.23$\pm$0.21*}}
& \phantom{0} 3.17\phantom{0} \\

 & & HR & \phantom{0}6.90$\pm$0.34
& \phantom{0}6.74$\pm$0.20
& \phantom{0}6.58$\pm$0.32
& \phantom{0}\textcolor{second}{\textbf{\underline{8.06$\pm$0.27}}}
& \phantom{0}7.17$\pm$0.20
& \phantom{0}7.47$\pm$0.39
& \phantom{0}7.78$\pm$0.39
& \phantom{0}7.96$\pm$0.61
& \phantom{0}7.92$\pm$0.38 
& \textcolor{first}{\textbf{\phantom{0}8.17$\pm$0.27*}}
& \phantom{0} 1.36\phantom{0} \\ \cline{2-14}
 & \multirow{2}{*}{Cold-start} & NDCG & \phantom{0}3.16$\pm$0.19
& \phantom{0}3.15$\pm$0.30
& \phantom{0}2.97$\pm$0.38
& \phantom{0}3.63$\pm$0.25
& \phantom{0}3.51$\pm$0.38
& \phantom{0}3.73$\pm$0.37
& \phantom{0}3.80$\pm$0.33
& \phantom{0}3.83$\pm$0.48
& \phantom{0}\textcolor{second}{\textbf{\underline{ 4.00$\pm$0.35}}}
& \textcolor{first}{\textbf{\phantom{0}4.39$\pm$0.21*}}
& \phantom{0} 9.75\phantom{0} \\
 & & HR & \phantom{0}6.26$\pm$0.39
& \phantom{0}6.11$\pm$0.42
& \phantom{0}6.03$\pm$0.45
& \phantom{0}7.40$\pm$0.39
& \phantom{0}7.18$\pm$0.51
& \phantom{0}7.18$\pm$0.74
& \phantom{0}7.56$\pm$0.56
& \phantom{0}\textcolor{second}{\textbf{\underline{7.94$\pm$0.85}}}
& \phantom{0}7.56$\pm$0.51 
& \textcolor{first}{\textbf{\phantom{0}8.63$\pm$0.19*}}
& \phantom{0} 8.69\phantom{0} \\ \cline{2-14}
 & \multirow{2}{*}{All} & NDCG
& \phantom{0}3.98$\pm$0.19
& \phantom{0}3.89$\pm$0.06
& \phantom{0}3.88$\pm$0.14
& \phantom{0}4.40$\pm$0.17
& \phantom{0}4.13$\pm$0.14
& \phantom{0}4.36$\pm$0.21
& \phantom{0}\textcolor{second}{\textbf{\underline{4.52$\pm$0.15}}}
& \phantom{0}4.49$\pm$0.29
& \phantom{0}4.46$\pm$0.19
& \textcolor{first}{\textbf{\phantom{0}5.79$\pm$0.29*}}
& \phantom{0}3.23\phantom{0} \\

 & & HR 
& \phantom{0}7.59$\pm$0.39
& \phantom{0}7.33$\pm$0.14
& \phantom{0}7.36$\pm$0.27
& \phantom{0}8.54$\pm$0.32
& \phantom{0}7.77$\pm$0.34
& \phantom{0}8.30$\pm$0.38
& \phantom{0}\textcolor{second}{\textbf{\underline{8.68$\pm$0.22}}}
& \phantom{0}8.54$\pm$0.50
& \phantom{0}8.66$\pm$0.29
& \textcolor{first}{\textbf{\phantom{0}10.61$\pm$0.22*}}
& \phantom{0}2.35\phantom{0}  \\ \hline

 \multirow{6}{*}{Elec} & \multirow{2}{*}{Tailed} & NDCG & \phantom{0}6.96$\pm$0.23
& \phantom{0}6.74$\pm$0.25
& \phantom{0}6.78$\pm$0.29
& \phantom{0}7.63$\pm$0.19
& \phantom{0}7.09$\pm$0.24
& \phantom{0}7.78$\pm$0.13
& \phantom{0}7.64$\pm$0.10
& \phantom{0}7.60$\pm$0.30
& \phantom{0}\textcolor{second}{\textbf{\underline{7.77$\pm$0.11}}}
& \textcolor{first}{\textbf{\phantom{0}8.00$\pm$0.10*}}
& \phantom{0} 2.96 \phantom{0} \\

 & & HR & \phantom{0}11.65$\pm$0.48
& \phantom{0}11.39$\pm$0.47
& \phantom{0}11.49$\pm$0.53
& \phantom{0}\textcolor{second}{\textbf{\underline{12.83$\pm$0.41}}}
& \phantom{0}11.66$\pm$0.56
& \phantom{0}13.05$\pm$0.28
& \phantom{0}12.45$\pm$0.25
& \phantom{0}12.56$\pm$0.39
& \phantom{0}12.66$\pm$0.28 
& \textcolor{first}{\textbf{\phantom{0}13.49$\pm$0.19*}}
& \phantom{0} 5.14\phantom{0} \\ \cline{2-14}
 & \multirow{2}{*}{Cold-start} & NDCG & \phantom{0}9.35$\pm$0.33
& \phantom{0}9.22$\pm$0.19
& \phantom{0}9.16$\pm$0.19
& \phantom{0}9.73$\pm$0.65
& \phantom{0}9.59$\pm$0.35
& \phantom{0}9.85$\pm$0.30
& \phantom{0}\textcolor{second}{\textbf{\underline{9.90$\pm$0.25}}}
& \phantom{0}9.76$\pm$0.48
& \phantom{0}9.73$\pm$0.40 
& \textcolor{first}{\textbf{\phantom{0}10.26$\pm$0.42*}}
& \phantom{0} 3.64\phantom{0} \\
 & & HR & \phantom{0}14.71$\pm$0.49
& \phantom{0}14.76$\pm$0.60
& \phantom{0}14.94$\pm$0.51
& \phantom{0}15.76$\pm$1.24
& \phantom{0}15.00$\pm$1.00
& \phantom{0}\textcolor{second}{\textbf{\underline{15.94$\pm$0.43}}}
& \phantom{0}15.24$\pm$0.60
& \phantom{0}15.35$\pm$0.90
& \phantom{0}15.53$\pm$0.34
& \textcolor{first}{\textbf{\phantom{0}17.12$\pm$0.39*}}
& \phantom{0} 7.40\phantom{0} \\ \cline{2-14}

& \multirow{2}{*}{All} & NDCG & \phantom{0}8.20$\pm$0.22
& \phantom{0}8.06$\pm$0.27
& \phantom{0}8.08$\pm$0.34
& \phantom{0}8.58$\pm$0.16
& \phantom{0}7.84$\pm$0.08
& \phantom{0}9.00$\pm$0.12
& \phantom{0}8.95$\pm$0.13
& \phantom{0}8.84$\pm$0.19
& \phantom{0}\textcolor{second}{\textbf{\underline{9.04$\pm$0.04}}}
& \textcolor{first}{\textbf{\phantom{0}9.33$\pm$0.09*}}
& \phantom{0} 3.22\phantom{0} \\

 & & HR & \phantom{0}13.08$\pm$0.43
& \phantom{0}12.81$\pm$0.45
& \phantom{0}12.86$\pm$0.59
& \phantom{0}13.60$\pm$0.50
& \phantom{0}12.31$\pm$0.19
& \phantom{0}\textcolor{second}{\textbf{\underline{14.46$\pm$0.31}}}
& \phantom{0}14.03$\pm$0.21
& \phantom{0}13.96$\pm$0.26 
& \phantom{0}14.08$\pm$0.26
& \textcolor{first}{\textbf{\phantom{0}15.28$\pm$0.26*}}
& \phantom{0} 5.64\phantom{0} \\

\midrule
\midrule

\multirow{6}{*}{Game} & \multirow{2}{*}{Tailed} & NDCG  
& \phantom{0}5.08$\pm$0.14
& \phantom{0}5.17$\pm$0.08
& \phantom{0}5.12$\pm$0.19
& \phantom{0}4.64$\pm$0.25
& \phantom{0}4.64$\pm$0.25
& \phantom{0}5.04$\pm$0.08
& \phantom{0}5.25$\pm$0.22 
& \phantom{0}\textcolor{second}{\textbf{\underline{5.31$\pm$0.15}}}
& \phantom{0}5.23$\pm$0.14
& \textcolor{first}{\textbf{\phantom{0}5.51$\pm$0.12*}}
& \phantom{0} 3.77 \phantom{0} \\

 & & HR 
& \phantom{0}9.48$\pm$0.38
& \phantom{0}9.46$\pm$0.14
& \phantom{0}9.39$\pm$0.34
& \phantom{0}8.45$\pm$0.38
& \phantom{0}8.45$\pm$0.38
& \phantom{0}9.34$\pm$0.17
& \phantom{0}9.73$\pm$0.35
& \phantom{0}\textcolor{second}{\textbf{\underline{9.95$\pm$0.39}}}
& \phantom{0}9.68$\pm$0.35
& \textcolor{first}{\textbf{\phantom{0}10.09$\pm$0.18*}}
& \phantom{0} 1.41 \phantom{0} \\ \cline{2-14}

 & \multirow{2}{*}{Cold-start} & NDCG 
& \phantom{0}9.13$\pm$0.64
& \phantom{0}\textcolor{second}{\textbf{\underline{9.97$\pm$1.37}}}
& \phantom{0}8.61$\pm$1.92
& \phantom{0}7.46$\pm$0.93
& \phantom{0}7.46$\pm$0.93
& \phantom{0}9.18$\pm$1.57
& \phantom{0}8.91$\pm$1.44
& \phantom{0}9.69$\pm$1.09
& \phantom{0}7.72$\pm$1.63
& \textcolor{first}{\textbf{\phantom{0}10.92$\pm$0.70*}}
& \phantom{0} 9.53\phantom{0} \\

 & & HR
& \phantom{0}20.00$\pm$3.95
& \phantom{0}19.15$\pm$1.90
& \phantom{0}16.60$\pm$2.48
& \phantom{0}15.32$\pm$1.59
& \phantom{0}15.32$\pm$1.59
& \phantom{0}16.17$\pm$2.89
& \phantom{0}16.60$\pm$2.48 
& \phantom{0}\textcolor{second}{\textbf{\underline{19.57$\pm$1.59}}}
& \phantom{0}15.74$\pm$2.89
& \textcolor{first}{\textbf{\phantom{0}22.55$\pm$2.17*}}
& \phantom{0} 15.20\phantom{0} \\ \cline{2-14}

 & \multirow{2}{*}{All} & NDCG 
& \phantom{0}5.39$\pm$0.15
& \phantom{0}5.40$\pm$0.09
& \phantom{0}\textcolor{second}{\textbf{\underline{5.57$\pm$0.10}}}
& \phantom{0}4.82$\pm$0.21
& \phantom{0}4.82$\pm$0.21
& \phantom{0}5.53$\pm$0.11
& \phantom{0}5.49$\pm$0.14 
& \phantom{0}5.51$\pm$0.09
& \phantom{0}5.30$\pm$0.05
& \textcolor{first}{\textbf{\phantom{0}5.79$\pm$0.12*}}
& \phantom{0}4.01\phantom{0} \\

 & & HR 

& \phantom{0}9.95$\pm$0.30
& \phantom{0}9.94$\pm$0.18
& \phantom{0}10.20$\pm$0.30
& \phantom{0}8.87$\pm$0.22
& \phantom{0}8.87$\pm$0.22
& \phantom{0}10.28$\pm$0.23
& \phantom{0}10.28$\pm$0.23
& \phantom{0}\textcolor{second}{\textbf{\underline{10.32$\pm$0.04}}}
& \phantom{0}9.92$\pm$0.19
& \textcolor{first}{\textbf{\phantom{0}10.61$\pm$0.15*}}
& \phantom{0}2.77\phantom{0} \\\hline

 \multirow{6}{*}{Video} & \multirow{2}{*}{Tailed} & NDCG
& \phantom{0}6.79$\pm$0.11
& \phantom{0}6.86$\pm$0.10
& \phantom{0}6.82$\pm$0.18
& \phantom{0}6.32$\pm$0.18
& \phantom{0}6.32$\pm$0.18
& \phantom{0}6.75$\pm$0.20
& \phantom{0}6.89$\pm$0.16
& \phantom{0}6.88$\pm$0.10
& \phantom{0}\textcolor{second}{\textbf{\underline{7.05$\pm$0.08}}}
& \textcolor{first}{\textbf{\phantom{0}7.08$\pm$0.14*}}
& \phantom{0} 0.43\phantom{0} \\
 
 & & HR  & \phantom{0}12.43$\pm$0.19
& \phantom{0}12.50$\pm$0.18
& \phantom{0}12.26$\pm$0.39
& \phantom{0}11.46$\pm$0.36
& \phantom{0}11.46$\pm$0.36
& \phantom{0}12.34$\pm$0.35
& \phantom{0}12.57$\pm$0.29
& \phantom{0}12.57$\pm$0.25
& \phantom{0}\textcolor{second}{\textbf{\underline{12.68$\pm$0.22}}}
& \textcolor{first}{\textbf{\phantom{0}12.95$\pm$0.26*}}
& \phantom{0} 2.13\phantom{0} \\ \cline{2-14}

 & \multirow{2}{*}{Cold-start} & NDCG
& \phantom{0}7.40$\pm$2.13
& \phantom{0}8.19$\pm$1.50
& \phantom{0}7.93$\pm$1.39
& \phantom{0}8.10$\pm$3.21
& \phantom{0}8.10$\pm$3.21
& \phantom{0}\textcolor{second}{\textbf{\underline{8.73$\pm$1.17}}}
& \phantom{0}8.57$\pm$1.31
& \phantom{0}8.54$\pm$2.13
& \phantom{0}8.26$\pm$1.71
& \textcolor{first}{\textbf{\phantom{0}9.62$\pm$1.58*}}
& \phantom{0} 10.10\phantom{0} \\

 & & HR 
& \phantom{0}18.67$\pm$2.67
& \phantom{0}21.78$\pm$3.27
& \phantom{0}20.00$\pm$2.43
& \phantom{0}17.78$\pm$4.66
& \phantom{0}17.78$\pm$4.66
& \phantom{0}21.78$\pm$2.59
& \phantom{0}\textcolor{second}{\textbf{\underline{22.22$\pm$2.81}}}
& \phantom{0}19.11$\pm$3.61
& \phantom{0}20.44$\pm$2.18
& \textcolor{first}{\textbf{\phantom{0}22.67$\pm$2.18*}}
& \phantom{0} 2.03\phantom{0} \\ \cline{2-14}

 & \multirow{2}{*}{All} & NDCG 
& \phantom{0}6.49$\pm$0.09
& \phantom{0}6.59$\pm$0.10
& \phantom{0}6.50$\pm$0.11
& \phantom{0}5.87$\pm$0.15
& \phantom{0}5.87$\pm$0.15
& \phantom{0}6.46$\pm$0.19
 & \phantom{0}6.04$\pm$0.19
& \phantom{0}6.58$\pm$0.16
& \phantom{0}\textcolor{second}{\textbf{\underline{6.67$\pm$0.11}}}
& \textcolor{first}{\textbf{\phantom{0}6.79$\pm$0.09*}}
& \phantom{0} 1.80\phantom{0} \\

 & & HR 
 & \phantom{0}12.10$\pm$0.16
& \phantom{0}12.18$\pm$0.14
& \phantom{0}11.97$\pm$0.31
& \phantom{0}10.92$\pm$0.22
& \phantom{0}10.92$\pm$0.22
& \phantom{0}12.14$\pm$0.25
& \phantom{0}11.25$\pm$0.30
& \phantom{0}12.32$\pm$0.29
&\phantom{0}\textcolor{second}{\textbf{\underline{12.33$\pm$0.16}}}
& \textcolor{first}{\textbf{\phantom{0}12.67$\pm$0.18*}}
& \phantom{0} 2.76\phantom{0} \\ 

\bottomrule
\end{tabular}
}
\begin{tablenotes}    
        \centering
\item[*]
{\footnotesize "*" denotes statistically significant improvements ($p < 0.05$), as determined by a paired t-test comparison with the second best result.}
\end{tablenotes}
\vspace{-10pt}
\end{table*}

\section{Model Analysis} 
\subsection{Analysis of Overlapping Ratio}
\label{sec:overlapping}
We designed experiments to test our model's performance with fewer cross-domain overlapping users by randomly downsampling the proportion of overlapping users (\(\mathcal{K}_{o}\)) in the training set while keeping the test set unchanged. We compared our model with SA-VAE at \(\mathcal{K}_{o}\) values of \{25\%, 50\%, 75\%, 100\%\}. This experiment simulates real-world scenarios, requiring the model to maintain stable performance with few overlapping users. Results are shown in Table \ref{compare_or}. As the number of overlapping users decreases, both SA-VAE and our model's performance decline due to the increased difficulty in capturing cross-domain signals. However, our model consistently outperforms SA-VAE at every \(\mathcal{K}_{o}\) level. This is because our model does not rely solely on overlapping users. While capturing cross-domain information becomes harder, our denoised pseudo-sequence effectively captures intra-domain interests. Additionally, the cross-domain and disentangle regularizers allow better inference on non-overlapping users for cross-domain signals based on behaviors in one domain.

\begin{table}[h!]
\small
\setlength\tabcolsep{0.3pt}
\setlength{\abovecaptionskip}{0pt}
\setlength{\belowcaptionskip}{1pt}
\centering

\caption{\centering Experiment results (\%) on "Cloth-Sport" dataset with the varying overlapping ratio ($\mathcal{K}_{o}$).}
\captionsetup{font=small}
\label{compare_or} 
\resizebox{0.5\columnwidth}{!}{
\begin{tabular}{c|c|c|cc|cc|cc|cc}
\toprule
\multirow{2}{*}{Domain} & \multirow{2}{*}{User} & \multirow{2}{*}{Metric} & \multicolumn{2}{c|}{$\mathcal{K}_{o}=25\%$ } & \multicolumn{2}{c|}{$\mathcal{K}_{o}=50\%$ } &\multicolumn{2}{c|}{$\mathcal{K}_{o}=75\%$} &\multicolumn{2}{c}{$\mathcal{K}_{o}=100\%$} \\ \cline{4-11}
 & & & SA-VAE\phantom{0} & i$^2$VAE & SA-VAE\phantom{0} & i$^2$VAE &  SA-VAE\phantom{0} & i$^2$VAE   &  SA-VAE\phantom{0} & i$^2$VAE\\ \midrule

\multirow{6}{*}{Cloth} & \multirow{2}{*}{Tailed} &NDCG
& \phantom{0}0.93	& \textcolor{first}{\textbf{\phantom{0}1.32}}	& \phantom{0}1.69	& \textcolor{first}{\textbf{\phantom{0}1.87}}	& \phantom{0}2.25	& \textcolor{first}{\textbf{\phantom{0}2.30}}
& \phantom{0} 2.40
& \textcolor{first}{\textbf{\phantom{0}2.52}\phantom{0}}\\

 & &HR
& \phantom{0}2.02	& \textcolor{first}{\textbf{\phantom{0}2.72}}	& \phantom{0}3.24	& \textcolor{first}{\textbf{\phantom{0}3.70}}	& \phantom{0}4.26	& \textcolor{first}{\textbf{\phantom{0}4.50}}
& \phantom{0}4.43
& \textcolor{first}{\textbf{\phantom{0}4.73}\phantom{0}}\\ \cline{2-11}

 & \multirow{2}{*}{Cold-Start} &NDCG
& \phantom{0}0.95	& \textcolor{first}{\textbf{\phantom{0}1.62}}	& \phantom{0}2.19	& \textcolor{first}{\textbf{\phantom{0}2.52}}	& \phantom{0}3.26	& \textcolor{first}{\textbf{\phantom{0}3.27}}
& \phantom{0}3.25
& \textcolor{first}{\textbf{\phantom{0}3.45}\phantom{0}}\\

 & &HR
& \phantom{0}2.07	& \textcolor{first}{\textbf{\phantom{0}3.39}}	& \phantom{0}4.83	& \textcolor{first}{\textbf{\phantom{0}5.20}}	& \phantom{0}6.27	& \textcolor{first}{\textbf{\phantom{0}6.21}}
& \phantom{0}5.89
& \textcolor{first}{\textbf{\phantom{0}6.77}\phantom{0}}
\\ \cline{2-11}

& \multirow{2}{*}{All} &NDCG
& \phantom{0}0.92	& \textcolor{first}{\textbf{\phantom{0}1.31}}	& \phantom{0}1.71	& \textcolor{first}{\textbf{\phantom{0}1.93}}	& \phantom{0}2.29	& \textcolor{first}{\textbf{\phantom{0}2.42}}
& \phantom{0} 2.52
& \textcolor{first}{\textbf{\phantom{0}2.59}}\phantom{0}\\

 & &HR
& \phantom{0}1.94	& \textcolor{first}{\textbf{\phantom{0}2.71}}	& \phantom{0}3.43	& \textcolor{first}{\textbf{\phantom{0}3.97}}	& \phantom{0}4.40	& \textcolor{first}{\textbf{\phantom{0}4.75}} & \phantom{0}4.81
& \textcolor{first}{\textbf{\phantom{0}5.00}}\phantom{0}\\ \hline

\multirow{6}{*}{Sport} & \multirow{2}{*}{Tailed} &NDCG
& \phantom{0}1.15	& \textcolor{first}{\textbf{\phantom{0}1.92}}	& \phantom{0}2.31	& \textcolor{first}{\textbf{\phantom{0}2.77}}	& \phantom{0}3.07	& \textcolor{first}{\textbf{\phantom{0}3.36}}
& \phantom{0}3.29
& \textcolor{first}{\textbf{\phantom{0}3.36}} \phantom{0}\\

 & &HR
& \phantom{0}2.48	& \textcolor{first}{\textbf{\phantom{0}4.06}}	& \phantom{0}4.59	& \textcolor{first}{\textbf{\phantom{0}5.60}}& \phantom{0}5.82	& \textcolor{first}{\textbf{\phantom{0}6.62}} 
& \phantom{0} 6.31
& \textcolor{first}{\textbf{\phantom{0}6.66}}\phantom{0}\\ \cline{2-11}

 & \multirow{2}{*}{Cold-Start} &NDCG
& \phantom{0}1.58	& \textcolor{first}{\textbf{\phantom{0}2.93}}	& \phantom{0}3.57	& \textcolor{first}{\textbf{\phantom{0}4.48}}	& \phantom{0}4.80	& \textcolor{first}{\textbf{\phantom{0}5.27}}
& \phantom{0} 5.05
& \textcolor{first}{\textbf{\phantom{0}5.43}}\phantom{0}\\

 & &HR
    & \phantom{0}3.37	& \textcolor{first}{\textbf{\phantom{0}6.13}}	& \phantom{0}6.80	& \textcolor{first}{\textbf{\phantom{0}8.82}}	& \phantom{0}9.29	& \textcolor{first}{\textbf{\phantom{0}10.51}}
& \phantom{0} 9.63
& \textcolor{first}{\textbf{\phantom{0}10.30}}\phantom{0}\\ \cline{2-11}

& \multirow{2}{*}{All} &NDCG
& \phantom{0}1.45	& \textcolor{first}{\textbf{\phantom{0}2.65}} 	& \phantom{0}3.08	& \textcolor{first}{\textbf{\phantom{0}3.85}} 	& \phantom{0}4.03	& \textcolor{first}{\textbf{\phantom{0}4.28} } 
& \phantom{0} 4.31
& \textcolor{first}{\textbf{\phantom{0}4.40}} \phantom{0}\\

 & &HR
& \phantom{0}2.97	& \textcolor{first}{\textbf{\phantom{0}5.35}} 	& \phantom{0}5.88	& \textcolor{first}{\textbf{\phantom{0}7.53}} 	& \phantom{0}7.65	& \textcolor{first}{\textbf{\phantom{0}8.22} } 
& \phantom{0} 8.06
& \textcolor{first}{\textbf{\phantom{0}8.53}} \phantom{0}\\
\bottomrule
\end{tabular}
}
\vspace{-12pt}
\end{table}

\subsection{Analysis of Parameter Sensitivity}
\label{sec:sensitivity}
\begin{figure}[h!]
\centering
\setlength{\abovecaptionskip}{0pt}
\setlength{\belowcaptionskip}{1pt}

\subfigure[Impact of $T$ on Cloth]{
\begin{minipage}[t]{0.27\linewidth}  
\centering
\includegraphics[width=0.99\linewidth]{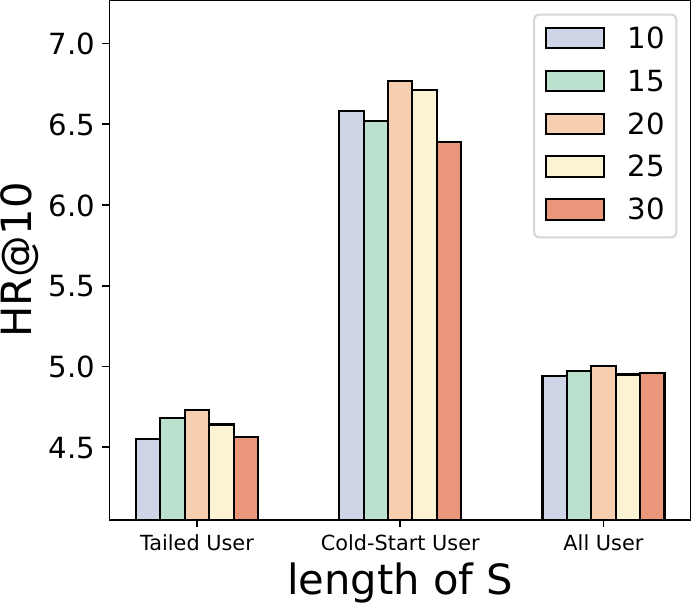}
\end{minipage}
}%
\subfigure[Impact of $\lambda_{a}$ on Cloth]{
\begin{minipage}[t]{0.27\linewidth}
\centering
\includegraphics[width=0.99\linewidth]{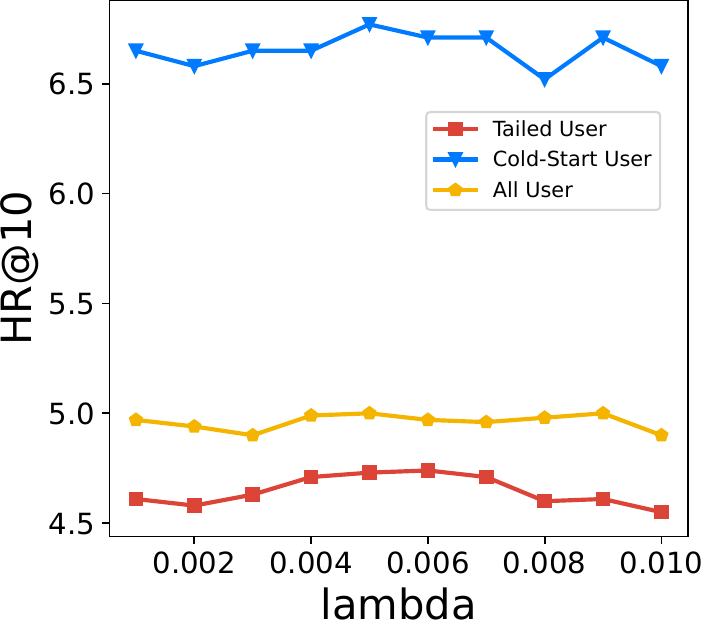}
\end{minipage}
}%
\subfigure[Impact of $\lambda_{c}$ on Cloth]{
\begin{minipage}[t]{0.27\linewidth}
\centering
\includegraphics[width=0.99\linewidth]{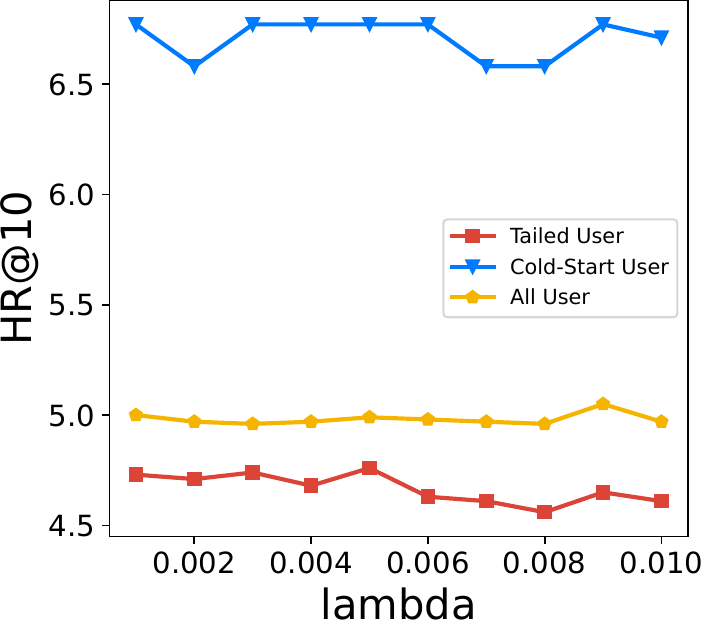}
\end{minipage}
}\\  

\subfigure[Impact of $T$ on Sport]{
\begin{minipage}[t]{0.27\linewidth}
\centering
\includegraphics[width=0.99\linewidth]{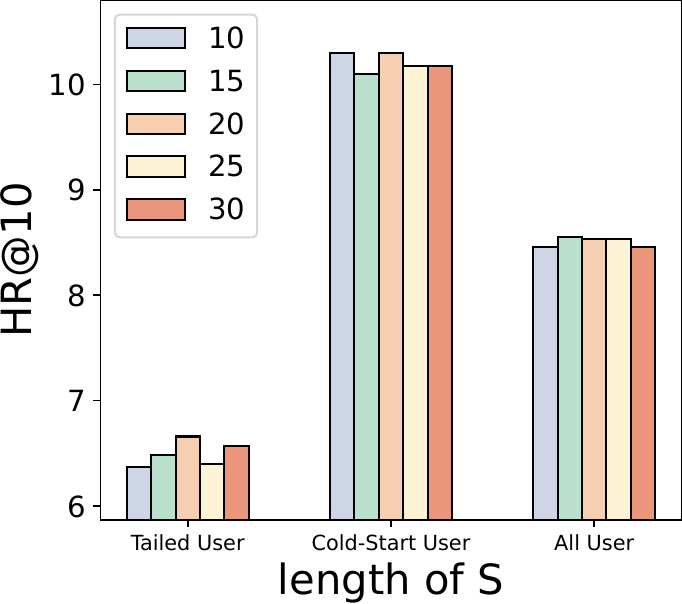}
\end{minipage}
}%
\subfigure[Impact of $\lambda_{a}$ on Sport]{
\begin{minipage}[t]{0.27\linewidth}
\centering
\includegraphics[width=0.99\linewidth]{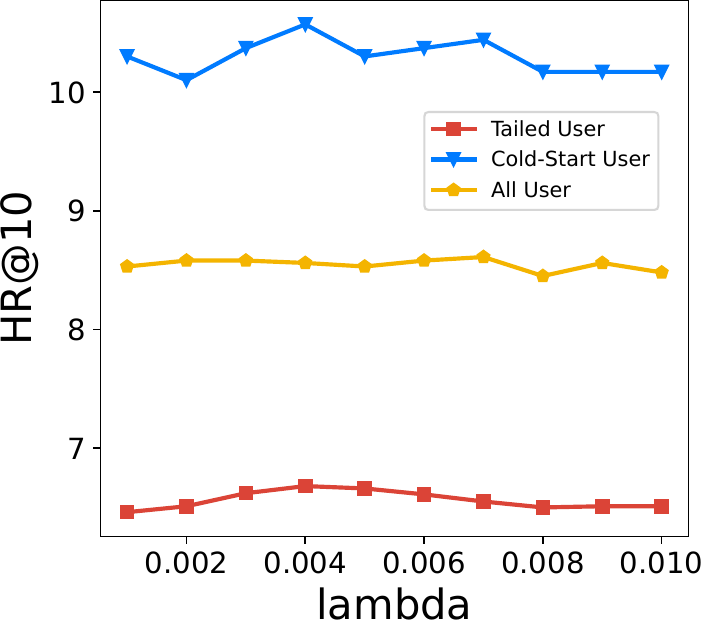}
\end{minipage}
}%
\subfigure[Impact of $\lambda_{c}$ on Sport]{
\begin{minipage}[t]{0.27\linewidth}
\centering
\includegraphics[width=0.99\linewidth]{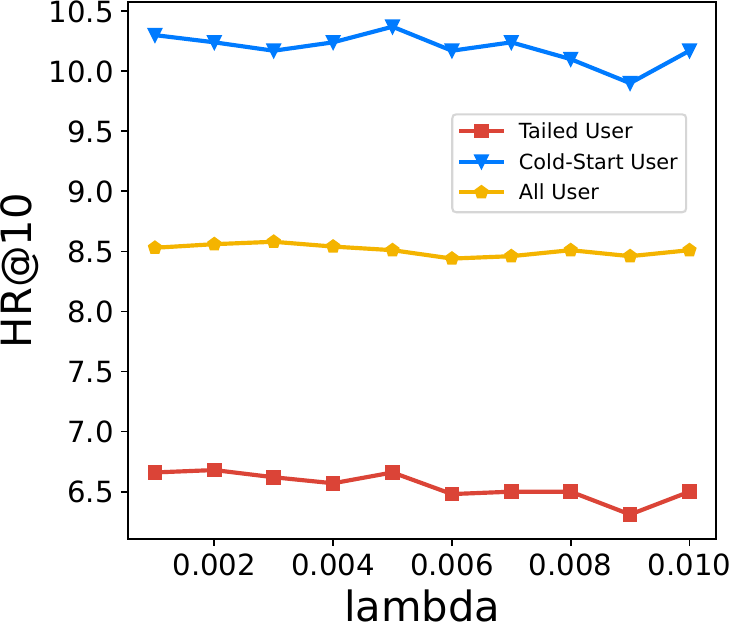}
\end{minipage}
}%

\caption{Experiment results (\%) of parameter sensitivity on the "Cloth" and "Sport" domains, respectively.}
\captionsetup{font=small}
\label{sensitivity}
\vspace{-12pt}
\end{figure}

In this section, we investigate the parameter sensitivity of sequence length \(T\) and the harmonic factors \(\lambda_a\) and \(\lambda_c\). Figures \ref{sensitivity}(a) and (d) show the impact of sequence length \(T\) on model performance (HR@10) in the Cloth and Sport domains, with \(T\) varying in \{10, 15, 20, 25, 30\}. The model performs best when \(T = 20\). Increasing \(T\) from 10 to 20 improves performance due to richer historical information, but performance decreases for \(T > 20\) due to padding items. Figures \ref{sensitivity}(b) and (e) show the impact of \(\lambda_a\) on model performance with \(\lambda_a\) varying in \{0.001, 0.002, ..., 0.010\} and \(\lambda_c\) fixed at 0.001. The model performs best when \(\lambda_a\) is around 0.004 or 0.005, indicating effective denoising without depressing the classification loss. Similarly, Figures \ref{sensitivity}(c) and (f) show the impact of \(\lambda_c\) on model performance with \(\lambda_a\) fixed at 0.005. The model achieves superior results when \(\lambda_c\) is set to 0.001 or 0.005.

\end{document}